\input epsf
\epsfverbosetrue
\documentclass[useAMS,usenatbib]{mn2e}
\usepackage{epsfig}
\title[Light and colour of LDN~1642]
{
{Light and colour of cirrus, translucent and opaque dust\\ 
in the high-latitude area of LDN\,1642 
}
\thanks{Based on observations collected at the European Organisation for Astronomical 
Research in the Southern Hemisphere; and based on observations with ISO, an ESA project with 
instruments funded by ESA member states(especially the PI countries: France, Germany, 
the Netherlands and the United Kingdom) and with the participation of ISAS and NASA.}
} 

\author[K. Mattila et al.]{K. Mattila$^{1}$\thanks{E-mail: kalevi.mattila@helsinki.fi}, 
P. V\"ais\"anen$^{2,3}$,
K. Lehtinen$^{1}$,
L. Haikala$^{4}$,
and M. Haas$^{5}$ 
\\
$^{1}$Department of Physics, University of Helsinki, P.O. Box 64, FI-00014 Helsinki, Finland\\
$^{2}$South African Astronomical Observatory, P.O. Box 9 Observatory, Cape Town, South Africa\\
$^{3}$Southern African Large Telescope, P.O. Box 9 Observatory, Cape Town, South Africa\\
$^{4}$Instituto de Astronom\'ia y Ciencias Planetarias, Universidad de Atacama, Copayapu 485, Copiapo, Chile\\
$^{5}$Astronomisches Institut, Ruhr-Universit\"at Bochum, Universit\"atsstrasse 150, D-44801 Bochum, Germany}

\begin{document}
\newcommand{\ana}{A\&A~}  
\newcommand{\apj}{{ApJ~}}     
\newcommand{\mnras}{{MNRAS~}} 
\newcommand{\ea}{{et~al.~}}
\newcommand{\cgs}{$10^{-9}$\,erg~cm$^{-2}$s$^{-1}$sr$^{-1}$\AA$^{-1}$}
\newcommand{\ang}{\AA \ }

\date{Accepted ; Received ; in original form }

\pagerange{\pageref{firstpage}--\pageref{lastpage}} \pubyear{2002}

\maketitle

\label{firstpage}

\begin{abstract}

{We have performed a 5-colour surface photometric study of the 
high-galactic-latitude area of dark nebula LDN\,1642. Scattered light properties are presented 
of diffuse, translucent and opaque dust over the range of 3500 -- 5500 \AA\,.
Far infrared absolute photometry at 200 $\mu$m improves 
the precision of and provides a zero point to the extinction. 
The intensity of the scattered light depends on dust column density in a characteristic
way: for optically thin dust the intensity first increases linearly, then turns to
a saturation value; at still larger extinctions the intensity
turns down to a slow decrease. The $A_V$ value of the saturated intensity maximum shifts
in a systematic way, from   $A_V\approx 1.5$\,mag at 3500 \AA\, to  $\sim 3$\,mag at 5500 \AA\,.
The intensity curves offer a straight-forward explanation for the behaviour of the 
scattered-light colours. At the intensity peak the colour  agrees with the integrated starlight colour,
while it is bluer at the low- and redder at the high-column-density side of the peak, respectively.
These colour changes are a direct consequence of the wavelength dependence of the extinction. 
We have compared the colours of the LDN\,1642 area with 
other relevant observational studies: high-latitude diffuse/translucent 
clouds, wide-field cirrus dust; and externally illuminated AGB-star envelopes. 
For extragalactic low-surface-brightness sources cirrus is an unwanted foreground contaminant. 
Our results for cirrus colours can help to distinguish cases where a diffuse plume or stream, 
apparently associated with a galaxy or a group or cluster, is more likely a local cirrus structure.
}
\end{abstract}


\section{Introduction}

With the growing interest in low-surface-brightness objects and phenomena in recent years
also the sky background brightness studies have gained new impetus \citep{iaus355}. The
dominant sky background components for an observer based in the inner Solar System or on the ground are
the zodiacal light and the airglow. In the outer solar system and beyond, 
the sky background is dominated by scattered light off the omnipresent dust particles. Clearly, 
it is important to know both the brightness and colour of this 'Galactic sky background'.  
Several aspects of the intensity of the scattered light have been analysed previously (see
for example \citealt{witt08, brandt, mat18}, and references therein).
In the present paper one of the main aims is to study besides intensity also the colour 
of the scattered light in cirrus, translucent and dense areas of interstellar dust.

The colour gives clues on the properties of the dust particles and the surrounding interstellar 
radiation field (ISRF) to which the dust particles are exposed. It is also important because 
at low surface brightness levels the question frequently arises what is the nature 
of a plume or loop of diffuse light seen in the outskirts of a galaxy, or towards the 'empty' 
intra-galactic space of a group or a cluster of galaxies (see e.g. \citealt{arp}). 
Is it an accumulation of unresolved stars stripped off by an interacting galaxy; 
or an enhanced-density clump in the diffuse intra-cluster medium? Alternatively, it may 
just be a local dust feature in the Solar neighbourhood. 
Difference of colour between extragalactic plumes and foreground dust clouds can be of help. 
Recent studies based on low-dust-opacity regions indicate that foreground cirrus exhibits
bluer colours as compared to galaxies and their associated structures.   \citet{roman} 
have discussed the colour differences between galaxies and foreground cirrus using $r-i$ and $i-z$ 
vs $g-r$ colours. \citet{rudick} and \citet{mihos17}, on the other hand, have detected 
in the Virgo cluster centre $B-V$ colour differences between intra-cluster light and foreground 
cirrus dust. Our study in the present paper makes use of both the $U-B$ and $B-V$ colours.
 As compared to the $r-i$ and $i-z$ vs $g-r$ diagrams, our $U-B$ vs $B-V$ analysis 
allows us to better identify the low-opacity cirrus structures; they show, especially,
a substantial blue excess of their $U-B$ colours. We will also study how the colours change
over different ranges of dust opacity, from cirrus to translucent and dense areas.

Instrumental progress has enabled the observations to penetrate ever deeper into the realm of 
the low-surface-brightness sky components. The CCD detectors have been the essential technical
aid, first of all for studies of individual objects or fields with a limited angular size.
Single-channel surface photometry with photomultiplier as detector has, however, 
still some advantages over the 
CCD imaging. The relatively small field of view of most telescope/CCD systems makes it hard to survey 
dark nebulae or other targets of large angular size. Deep surface photometry is also made difficult 
because accurate flat-fielding is needed over degree-scales. 
Special methods, appropriate for accurate low-surface-brightness photometry over large field sizes, 
have been utilized in those cases. A thorough discussion of our methods is presented 
in Appendix A of this paper.  

In the present paper we will analyse the surface-photometric results in the high-galactic-latitude 
area around LDN~1642. 
Our five-colour intermediate-band photometry covers  the range 3500 - 5500 \AA\,.
An important ingredient of the present analysis is that 200 $\mu$m absolute photometry 
is available for all our 
positions. 
These data provide a zero point for the dust column densities in our target area and, in addition, 
considerably improve the precision of the extinction estimates.

LDN\,1642 has been chosen for our detailed surface photometric study mainly for two reasons:
{\em firstly}, its surrounding sky has exceptionally low extinction,
 $A_V \sim 0.1 - 0.2$\,mag, which means that the scattered light contribution is low there.
 LDN~1642 and its 
surroundings provide a representative test area to study the scattered light 
from dust over a wide range, covering areas from diffuse/cirrus dust over translucent  
up to very high extinction areas.
{\em Secondly}, the dense core of the cloud with  $A_V \sim 16$\,mag,
is expected to provide substantial obscuration of the extragalactic background light.
 
A dark nebula acts as a foreground screen that blocks the light along the line of sight (LOS) 
from stars and the background sky, { including also}
the extragalactic background light (EBL).
The attenuation of the EBL 
is strong towards the opaque central core while very little shadowing occurs 
in the transparent outer areas. 
Because the intensity of the  the scattered starlight is higher than that of the EBL, 
the EBL has only a minor effect on the observed surface brightnesses. 
Previously, we have used medium-resolution spectroscopic observations of the surface brightness 
of LDN\,1642 for a measurement { of} the EBL. This approach has enabled the scattered-light subtraction
by means of the depths of strong Fraunhofer lines and the 4000 \AA\, discontinuity in the 
integrated starlight (ISL) spectrum 
\citep{mat17b,mat19}. Based on these and other current results, the EBL would, however, 
make only a marginal effect on intermediate-band photometric 
observations, such as for LDN\,1642 as in the present paper.  
 
This paper is composed as follows. The observational data are briefly described
in Section 2, while more details on methods and  data reductions are given in Appendices A and B.
In Section 3 we present the basic analysis and results, the extinction dependence of the intensity, colour 
and SED of the scattered light, from diffuse over translucent to opaque dust.
In Section 4 we first discuss the reasons for the colour and SED variations in terms
of the SED of the incident radiation field and the illumination geometry. 
We then compare our results for the LDN\,1642 area with the colours of a cirrus/translucent 
cloud sample and some circumstellar AGB-star envelopes. 
Finally we ask whether the colours could be used to tell whether some of the diffuse plumes,
apparently associated with a galaxy or a cluster, might actually be foreground cirrus features.
 
\begin{figure} 
\vspace{-10pt} 
  \hspace{-0cm}
\includegraphics[width=75mm, angle = 0]{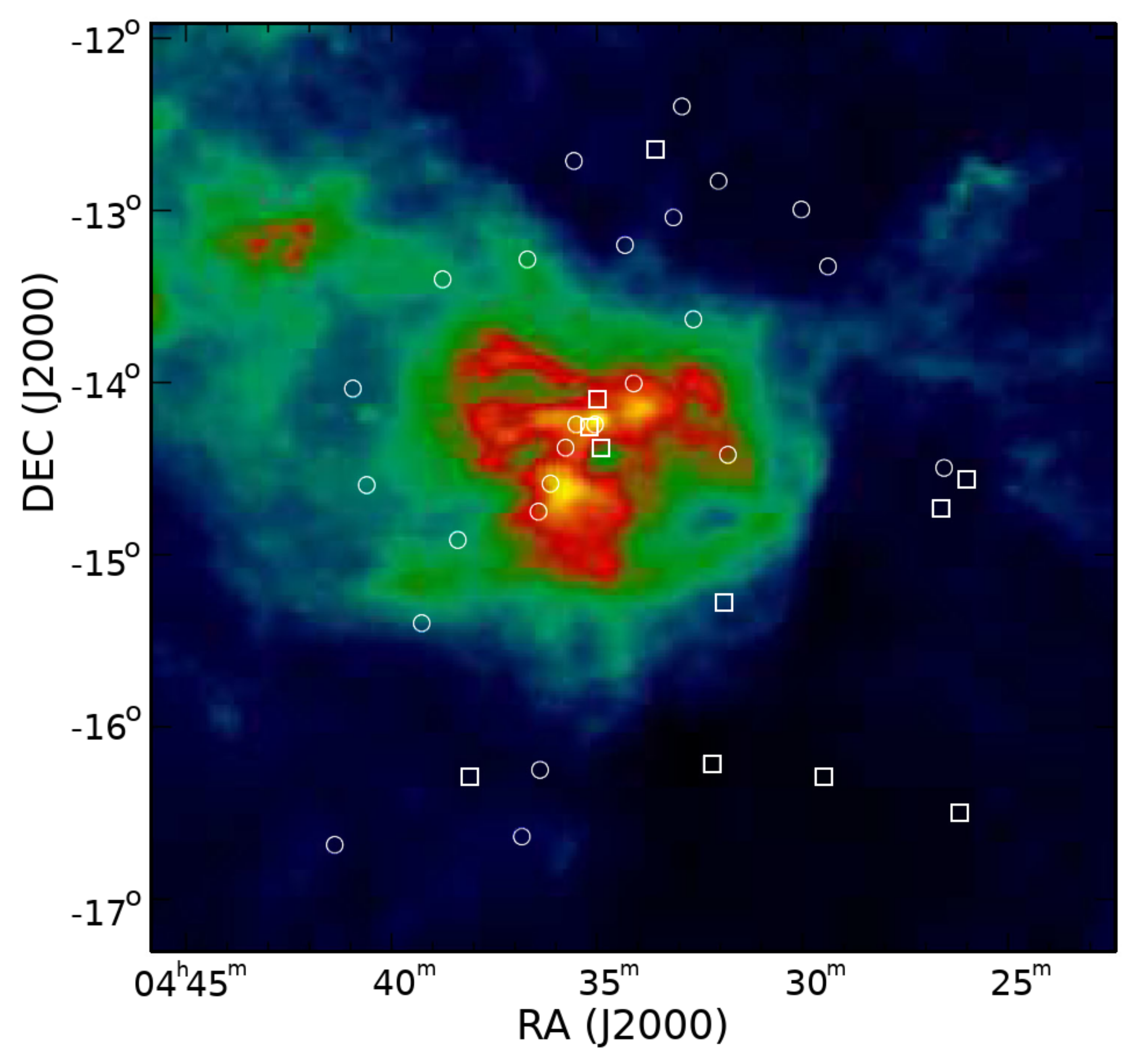}
\vspace{-0.0cm}
\hspace{0.5cm}
\includegraphics[width=75mm, angle = -0]{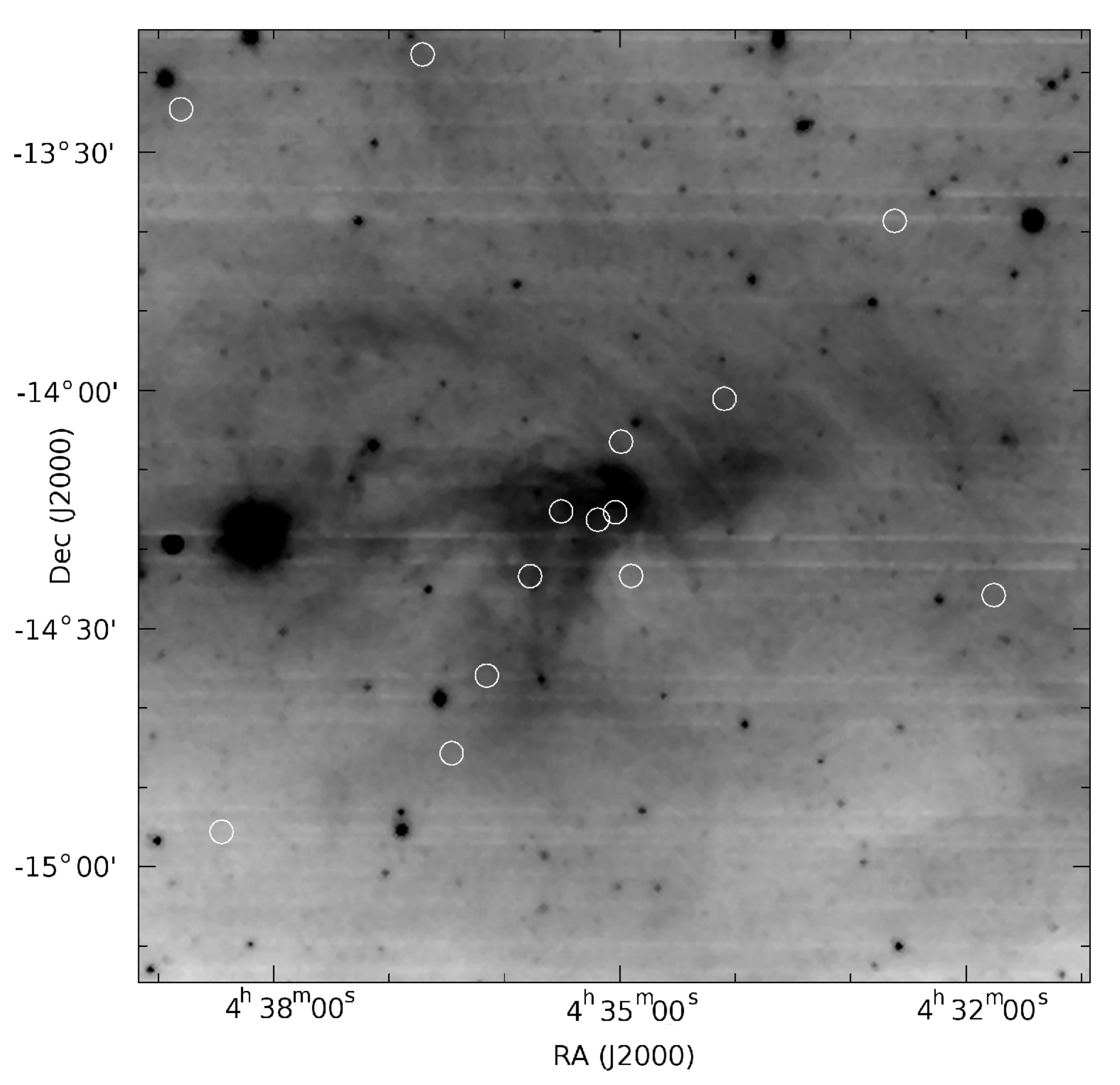}
\vspace{0.9 cm}
\hspace{1.2cm}
\includegraphics[width=67mm, angle = -0]{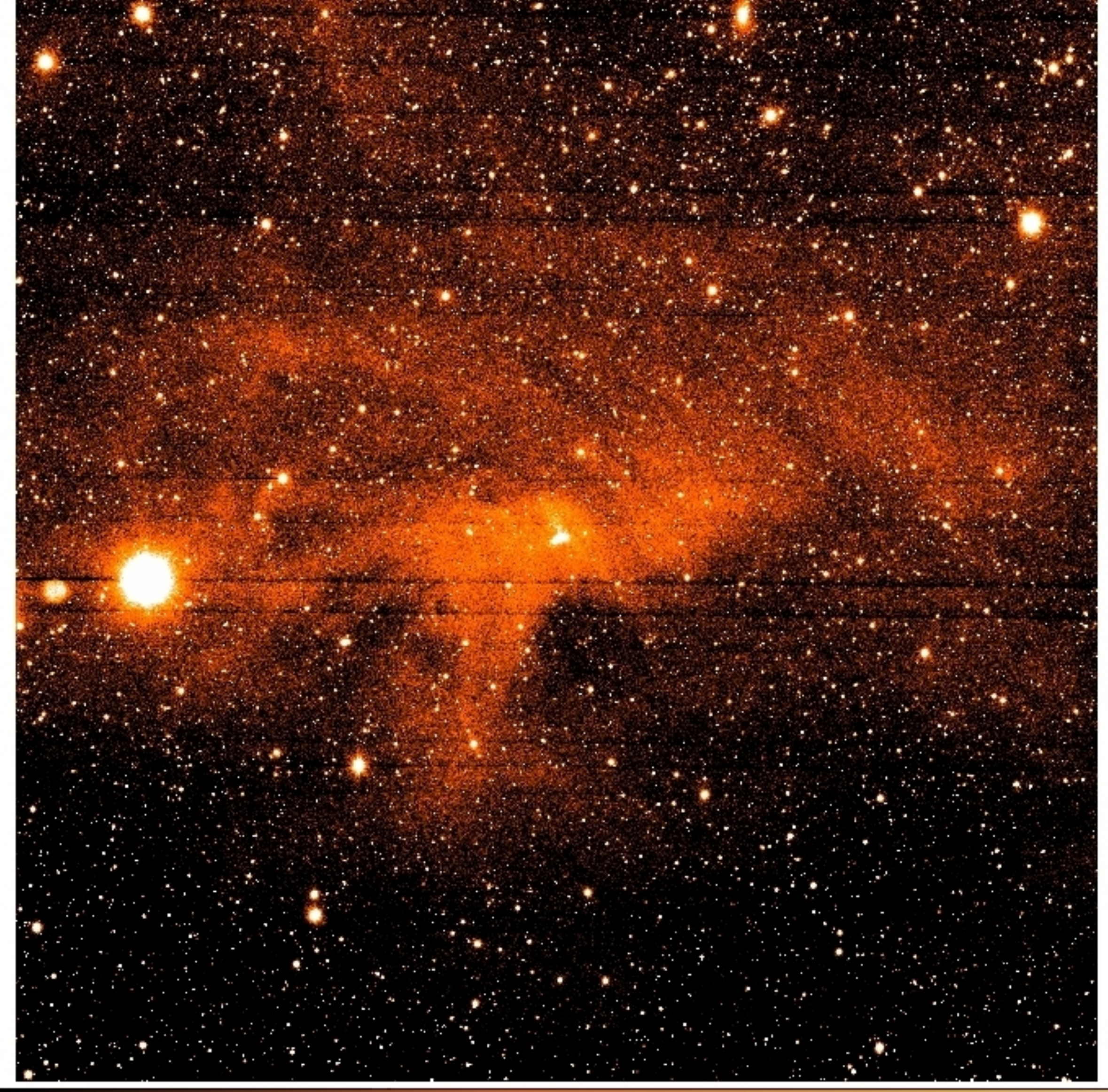}
\vspace{-1cm}
\hspace{2.6 cm}
\caption{{\em Upper panel.} The observed positions in LDN\,1642 cloud area superimposed on
a map of IRAS 100~$\mu$m emission that serves as a qualitative proxy for the interstellar 
extinction through the cloud. { Squares indicate positions used also in
the spectroscopic observations by \citet{mat17a,mat17b}}. 
{\em Middle panel.} Locations of the photometric positions 
superimposed on a VYSOS $i$ band (7480 \AA) image of the central $\sim 2 \times 2$\,deg area 
of LDN\,1642. Faint stars as seen in the original  $i$ band image in the {\em bottom panel} 
have been removed. 
}
\vspace{-10pt}
\label{L1642_image}
\end{figure}

\begin{figure} 
\vspace{-30pt} 
\includegraphics[width=70mm, angle = 0]{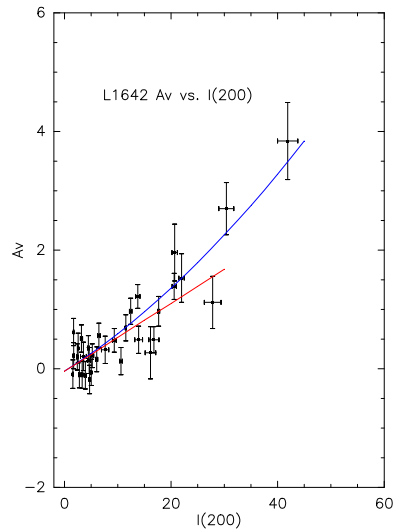}
\vspace{-20pt}
\caption{Visual extinction from 2MASS $JHK$ photometry vs $I_{\rm 200}$ from absolute ISOPHOT 
photometry for 35 positions 
in the LDN\,1642 area. A second-order polynomial fit to the positions at  
$I_{\rm 200} < 45$ MJy/sr is shown as blue line. The red line is a linear fit to the values at 
$I_{\rm 200} < 28$ MJy/sr. }
\vspace{-00pt}
\label{AVvsI200}
\end{figure}

\section{The data used for the analysis}

\subsection{Optical surface brightness data}
Our five-colour surface photometry covers an area of $\sim 4\degr \times 4\fdg5$ centered on 
LDN~1642. Within this area 35 selected positions with aperture $\O$$88\arcsec$, free of stars down to 
$B\sim20$\,mag, were measured 
differentially relative to a standard reference position in the centre of the cloud. 
The area includes a wide range
of dust environments, from the high opacity core with $A_V \approx 16$\,mag over translucent regions
($A_V \sim 1 - 5$\,mag) to the cirrus-like outskirts with  $A_V \sim 0.1 - 0.5$\,mag.
Aided by its location at high galactic latitude, $b = -36.5$\,deg, and seen towards the anticentre, 
$l = 211$\,deg, and at a $|z|$-distance of $\sim$65-100 pc \citep{hearty}, the surroundings  
 of LDN~1642 are relatively free of dust. The coordinates of the observed positions 
are given in  table {\small{\sc L1642\_AV+5colour+titleNEW.txt}} and are shown in Fig.\,1. 
(For the table, see Supporting information; for the VYSOS $i$ band image, see \citet{mat18} Section 2.3 and Fig.\,3).

The observations were 
carried out in five intermediate band filters, namely the standard $u$ (3500\AA\,), $b$ (4670 \AA\,) 
and $y$ (5550 \AA\,) Str\"omgren filters and two custom-made ones with centre wavelengths and FWHPs 
of 3840 \AA\,(175\AA\,) and 4160 \AA\,(140\AA\,). The latter two were tailored
for measurement of the 4000 \AA\, jump. 
The ESO 1-m and 50-cm telescopes at La Silla were used in parallel simultaneous observations;
both telescopes were equipped with closely identical filters; the 50-cm telescope was used for monitoring
the airglow variations.

The foreground surface brightness components were differentially removed: 
instrumental stray light from stars, 
the spatial gradients of airglow, tropospheric scattered light and zodiacal light. 
In addition, the emission from ionized gas was estimated and subtracted using H$\beta$ 
line measurements. 
Using the 200 $\mu$m emission intensities  $I_{200}$
the surface brightnesses were referred to a zero background intensity level in the following way: 
In each filter band the background intensity level was determined by means of a 3-dimensional fit 
where besides the coordinates R.A. and Dec. the third variable was  $I_{200}$. 
At small and moderate opacities $I_{200}$ is proportional to the dust column density; 
thus, by extrapolation to  $I_{200}$ = 0 we were able to
set the optical surface brightnesses to a system where  their zero level corresponds 
to zero dust column density. For a detailed description of the surface brightness observations and
reductions, see \citet{Vai94} and Appendix A.\footnote{The data reductions and analysis are largely based on the
MSc thesis of P. V\"ais\"anen \citep{Vai94}}

\begin{figure} 
\vspace{-30pt}
\hspace{-1cm} 
\includegraphics[width=125mm, angle = 0]{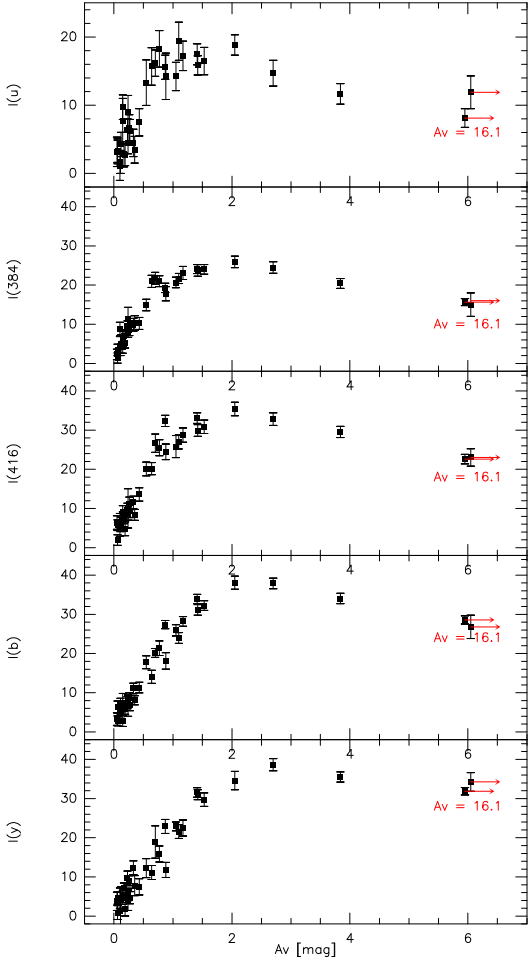}
\caption{Surface brightness in dependence of visual extinction $A_V$ for five optical bands 
from $u$ to $y$ in the LDN\,1642 area. Surface brightness unit is \cgs\,. Values for two adjacent
positions in the high extinction core with $A_V = 16.1$ are shown displaced to $A_V = 6$\,mag. }
\vspace{-0pt}
\label{Isca_vsAV}
\end{figure}

\begin{figure} 
\vspace{-30pt}
\hspace{-1cm} 
\includegraphics[width=120mm, angle = 0]{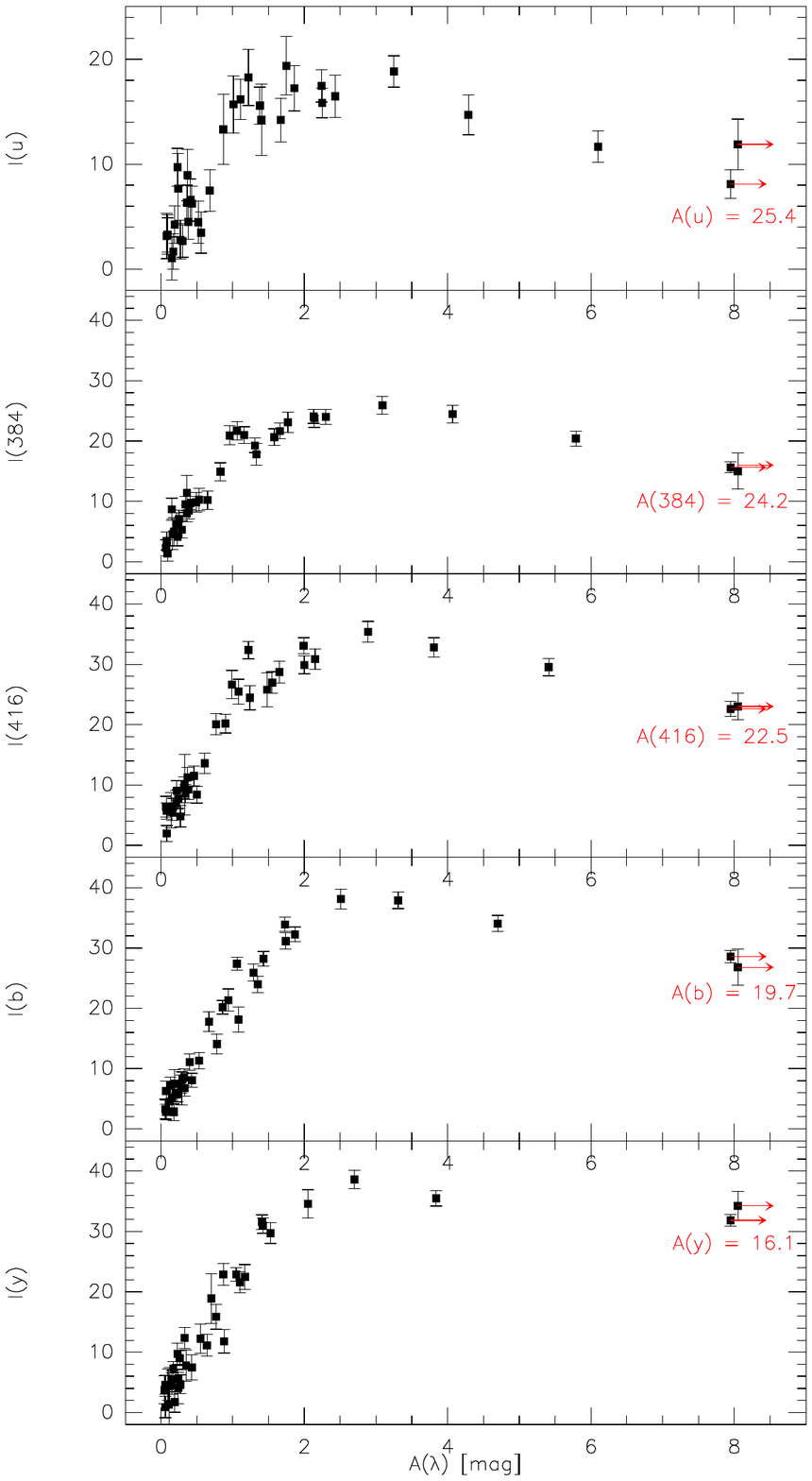}
\caption{Surface brightness in dependence of extinction $A({\lambda})$ for five optical bands 
from $u$ to $y$ in the LDN\,1642 area. Surface brightness unit is \cgs\,. For each band, $A({\lambda})$
is the extinction valid for the central wavelength of the band. It has been scaled from $A_V$ according to
the \citet{cardelli89} extinction law for  $R_V = 3.1$. Values for two adjacent
positions in the high extinction core with $A_V = 16.1$ are shown displaced to $A(\lambda) = 8$\,mag. }
\vspace{-0pt}
\label{Isca_vsAlamda}
\end{figure}

\vspace{20pt}

\subsection{Optical extinction from $JHK$ photometry}
Using  2MASS $JHK$ colour excesses of stars and the {\sc nicer} method \citep{lombardi} 
we have derived  extinctions  $A_V$ for most of our positions, see \citet{Lehtinen07}.  
In the outer areas ($r \ge 30\arcmin$ from centre) the extinctions were derived by using 
a $\O$$6\arcmin$ FWHM Gaussian as weighting function for the extinction values of individual
stars. In the inner areas ($ r < 30\arcmin$ from centre) where the extinction may 
change substantially over  $6\arcmin$, $\O$$3\arcmin$ FWHM was adopted. 
Because of the relatively low number density of 2MASS stars in the direction of L\,1642 
the extinction values had substantial statistical errors,   
ca. $\pm0.25$~mag and  $\pm0.45$~mag for $\O$$6\arcmin$ and  $\O$$3\arcmin$ FWHM, respectively.

The two highest opacity positions in the centre of the cloud have no 2MASS
stars within  $1\arcmin$  from the position centre. For these two positions 
we have made use of $H$ and $K_s$ band photometry with ESO's NTT/SOFI instrument.
The mean extinction in a $2\arcmin \times 4\arcmin$ area encompassing the two positions
was found to be {$A_{\rm V} = 16.1\pm1.4$\,mag}. A detailed description is given in Appendix B.

\subsection{Extinction and its zero point from 200 $\mu$m absolute photometry}

The ISOPHOT instrument \citep{lemke} aboard the $ISO$ satellite  \citep{kessler} was used 
to observe the surface brightness positions in and around LDN~1642 in the 
{\em absolute photometry mode} at 200~$\mu$m.
The zero point of $I_{200}$ was corrected for a zodiacal emission (ZE) intensity of 
$0.8\pm0.2$~MJy\,sr$^{-1}$ and a cosmic infrared background (CIB) of $1.1\pm0.3$~MJy\,sr$^{-1}$ 
\citep{hau}. At time of the 200~$\mu$m observations (1998-03-19/20) the 
longitude difference was $\lambda$(LDN\,1642)-$\lambda_{\sun}=63\fdg 7$; the ZE intensity 
at 200~$\mu$m was estimated using a 270 K blackbody fit to the ZE intensities at 100, 140 and 
240~$\mu$m. They were interpolated from the weekly DIRBE Sky and Zodi Atlas 
(DSZA)\footnote{https://lambda.gsfc.nasa.gov/product/.../dirbe\_dsza\_data\_get.cfm}
maps based on the \citet{kelsall} interplanetary dust distribution model.
The sum of these two corrections amounted to  $2.0$~MJy\,sr$^{-1}$, in agreement with 
the value based on extrapolation of the $I_{200}$ vs H{\sc i} 21-cm relationship to 
zero  H{\sc i} intensity (for the method see \citealt{juvela}). 

For low-to-moderate column densities the 200~$\mu$m intensities are well correlated with the 
optical extinctions,  $A_{\rm V}$, (see e.g. \citealt{Lehtinen07}).
A linear fit of $A_{\rm V}$ values from 2MASS JHK photometry vs  $I_{200}$ gave at 
$I_{200}\la $28 MJy\,sr$^{-1}$ the relationship (see Fig.~2).
\[
A_{\rm V} = (0.0574\pm 0.0079)(I_{200}/{\rm MJy\,sr}^{-1}) - 0.044\pm 0.092 \]

{ In spite of the substantial uncertainties of the 2MASS-based $A_{\rm V}$ values the slope and 
the zero point of the relationship are obtained with good precision.
The precision  of  of the $I_{200}$ values has been estimated using the intrinsic errors 
given by the ISOPHOT Interactive Analysis programme PIA \citep{gabriel} on the one, and the differences
between adjacent low-intensity points in the LDN~L1642 area on the other hand. These error estimates
amounted to $\sim\pm0.2 - \pm1 {\rm MJy\,sr}^{-1}$ for each of the two methods. 
The above relation enables, thus, a precision of $\sim\pm0.06$\,mag $(1\sigma)$ to be reached for 
the individual $A_{\rm V}$ values;} this is clearly better than that reached with 2MASS.
Therefore, visual extinctions for our low-to-moderate extinction positions, $A_V\la2$mag, 
were determined using the 200~$\mu$m observations as an intermediate step.
 
Because the zero point of the  200~$\mu$m intensities has been corrected for ZE 
and CIB, also the resulting extinctions  $A_{\rm V}$ are free of these offsets and refer to
the dust extinction only. { The uncertainty of the CIB+ZE correction of   
$\sim \pm0.5$  MJy\,sr$^{-1}$ and  of the zero-level term of the $A_{\rm V}$ vs. $I_{200}$ relationship 
of $\pm 0.092$\,mag introduce a systematic zero-point error of  $\sim \pm 0.12$\,mag
to the whole set of  $A_{\rm V}$ values.}

For the positions with $I_{200}\ge $30 MJy\,sr$^{-1}$ ( $A_{\rm V} \ga 2$mag) the 
points clearly deviate from the linear relationship (see Fig.~2). This is a well-known effect caused by
the decreasing heating power of the UV-optical-NIR radiation
towards the inner parts of an optically thick cloud (see e.g. \citealt{Lehtinen07}). 
For these positions
we have adopted the extinction values as they resulted from the 2MASS $JHK$ 
and NTT/SOFI  $HK$ colour excesses of stars. \\

The observational material used in this analysis consists of the following data for 35 positions in the LND\,1642 area: 
(1) differential surface brightnesses in the five optical filter bands plus absolute surface 
brightness at FIR 200 $\mu$m, see table 
{\sc Final2020JUL9N.txt}; \\
(2) coordinates of the positions and extinction values for the five filter bands, see table\\ 
{\small{\sc L1642\_AV+5colour+titleNEW.txt}}.\\
The tables are given in Supplementary information.
\noindent For the locations of the 35 positions, see also Fig.\,1.

\hspace{2cm}
\begin{figure*} 
\hspace{-1cm} 
\vspace{-110pt}
\includegraphics[width=150mm, angle = -90]{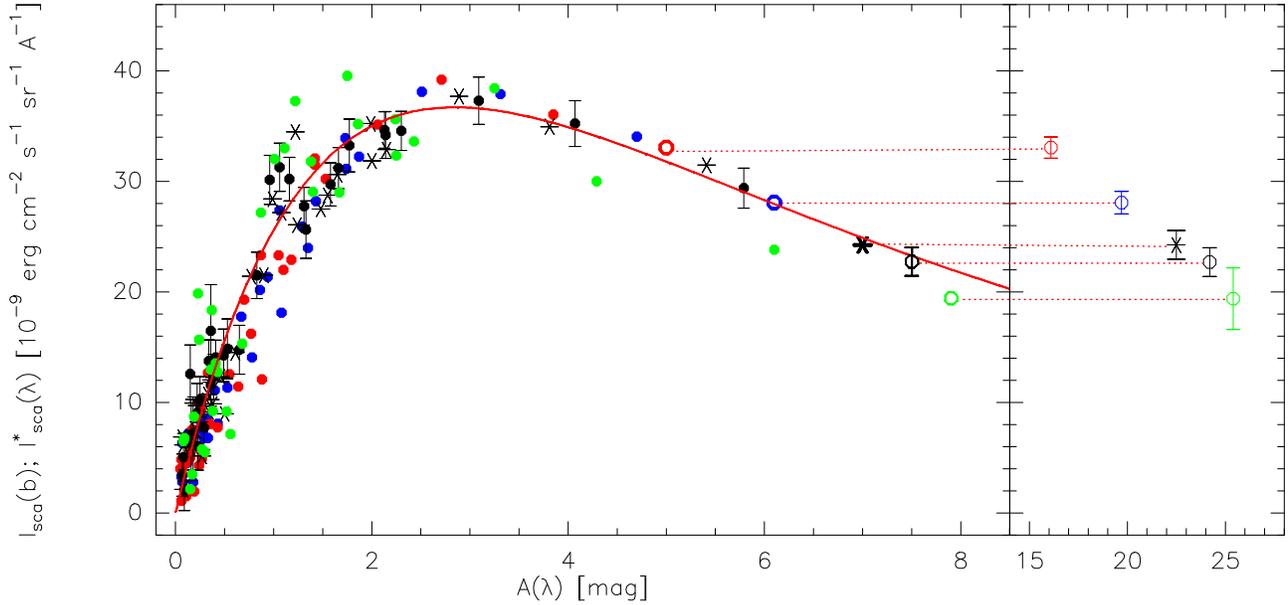}
\caption{Demonstration of the similarity of the observed $I_{\rm sca}(\lambda)$ vs $A({\lambda})$ 
relationships in the five optical bands. The intensity scale is for the $b$ band in units \cgs\,. 
For the other bands the surface brightness,  $I^*_{\rm sca}(\lambda)$, has been scaled 
by a factor which gives optimum agreement between the bands.
Horizontal axis is the extinction $A({\lambda})$, valid for each of the five bands.
It has been scaled according to  the \citet{cardelli89} extinction law for  $R_V = 3.1$.
The symbols and multiplication factors for the five bands are: $u$(green, 1.997), 384 nm (black dots 
and circles with error bars, 1.378), 416 nm (black asterisks, 1.041), $b$(blue, 1.000), $y$(red, 0.992).
The values for dark core position are shown in the separate box with  $14 < A_{\lambda} < 28$\,mag.
They are also shown replaced to extinctions $5 < A(\lambda) < 8$\,mag that correspond to their 
estimated {\em 'scattering-equivalent'} extinctions. For the two adjacent positions in the high-extinction 
core $A_{V} = 16.1$\,mag, the weighted mean values are shown. The red curve represents a {\tt mpfitfun} 
fit to the 384 nm data points. See text for clarifications.}
\vspace{0pt}
\label{Isca_vs_A_comb}
\end{figure*}

 
\section{Analysis and results}

\subsection{Scattered light intensity vs extinction}

In Fig.\,\ref{Isca_vsAV} the scattered light intensities, $I_{sca}(\lambda)$, for the five filter bands 
from $u$ to $y$ are shown as function of $A_V$. In Fig.\,\ref{Isca_vsAlamda} 
 values of $I_{sca}(\lambda)$ are shown as function of $A(\lambda)$, 
that is the extinction for the wave band in question. 
By using the FIR absolute photometry at 200 $\mu$m and its zero point we have been able to set the zero 
point also for the optical extinction,  $A_V$ and  $A(\lambda)$. 

For the darkest background positions in the outermost periphery of LDN~1642 
 the intensity range of $I_{200}$ is 1.7 to 5.2\,MJy\,sr$^{-1}$.
Applying the $A_V$ vs $I_{200}$ relationship the mean value of extinction for these 16 positions 
is  $A_V = 0.17\pm0.08$~mag or  $A(b) = 0.20\pm0.10$~mag; correspondingly, the mean of the scattered light 
intensities is  $I_{sca}(b)= (5.9\pm1.5)$\,\cgs\,. 
Although the observed surface photometry values are {\em differential}, representing
the difference between on-cloud and off-cloud positions, the intensities
in Figs.\,\ref{Isca_vsAV} and \ref{Isca_vsAlamda} are {\em absolute values relative
to zero dust column density} as set by the 200 $\mu$m absolute photometry. 

As shown by Figs.\,\ref{Isca_vsAV} and \ref{Isca_vsAlamda} the general shapes of the $I_{sca}$ 
vs extinction curves are very similar in all five bands, both for $A_V$ and  
 $A(\lambda)$ as x-axis. There are, however, significant differences, as well. 
In Fig.\,\ref{Isca_vsAV} one sees that the maximum of the curve shifts in a 
systematic way from $A_V \approx 1.5$ to $\approx 3$\,mag  when passing from the ultraviolet $u$ 
to the yellow $y$ band. The other parts of the curve  are shifted correspondingly. 

  On the other hand, when using  for each band its 'own' 
extinction, $A(\lambda)$,  the maximum occurs at closely the same x-axis position, 
$A(\lambda)\approx$ 2.5-3\,mag (Fig.\,\ref{Isca_vsAlamda}).
This agreement between the different bands is even more
clearly demonstrated by Fig.\,\ref{Isca_vs_A_comb} where the data points $I_{sca}(\lambda)$ vs  
$A(\lambda)$ have been overplotted in the same diagram. Here, the intensities have been 
adjusted by scaling factors so that an optimum agreement is obtained. These scaling factors mainly
reflect the band-to-band differences of the illumination source, the ISRF of starlight in the Solar 
neighbourhood. To a lesser degree they also reflect the wavelength dependence of  dust albedo;
this is, however, only a smaller effect as compared to the ISRF. { The dominance of the ISRF 
for the band-to-band differences is also demonstrated by the values of the scaling factors:
there is an almost perfect agreement of the (inverse) scaling factors, 0.501 ($u$), 0.726 (384 nm), 
0.961 (416 nm), 1.000 ($b$) and 1.008 ($y$), with the ISRF model spectrum; see
 Fig.~\ref{I_vslam_norm} {\em lower panel.} }

Considering the low surface brightnesses we are here dealing with, 
the precision of our  $I_{sca}$ vs  $A_V $ and  $I_{sca}$ vs  $A(\lambda)$ curves is remarkably good.
This is demonstrated both by the good position-to-position agreement of the individual curves as 
shown in Figs.\,\ref{Isca_vsAV} and \ref{Isca_vsAlamda} as well as by the good band-to-band
agreement as seen in Fig.\,\ref{Isca_vs_A_comb}. 

The shape of the  $I_{sca}$ vs extinction curve is well understood.
For $A(\lambda)\la 6$ it can be approximated by the {\em ansatz}:
\begin{equation}
I_{\rm sca}({\lambda})= C_{\rm ISRF}(\lambda)\,\, a_{\lambda}(1 - e^{-b_{\lambda}\tau(\lambda)})e^{-c_{\lambda}\tau(\lambda)}
\label{Eq1}
\end{equation}
where $C_{\rm ISRF}({\lambda})$ is a constant determined by the strength of the ambient starlight radiation 
field and  $a_{\lambda}$ is the albedo of the dust; the LOS optical depth is given by 
$\tau(\lambda)= 0.921\,A(\lambda)= 0.921 \frac{k_{\lambda}}{k_V} A_V$, where the extinction 
coefficients $k_{\lambda}$ are according to the \citet{cardelli89} extinction law for $R_V = 3.1$.
The second term $1 - e^{-b_{\lambda}\tau(\lambda)}$ describes the optically thin and the saturated
part of the curve up to  $A(\lambda)\la 2 -3$\,mag.
The third term, $e^{-c_{\lambda}\tau(\lambda)}$, becomes effective only at larger optical depths, 
$\tau(\lambda) \ga 3$, and describes the slow decline beyond the maximum of the $I_{\rm sca}({\lambda})$ 
vs  $A(\lambda)$ curve.

{\em 'Scattering-equivalent' extinctions.} When modelling the scattered light intensity at low and 
moderate opacities, $A(V) \la 6$,  the same opacity accounts for both extinction and 
scattering, that is $A^{\rm sca}(\lambda) \approx  A^{\rm ext}(\lambda)$. 
This is not valid for the highest opacity positions in the core, however.
In that case, the scattering is not determined by the total extinction through the dense core but 
by the dust shell around it,  and $A^{\rm sca}(\lambda) <  A^{\rm ext}(\lambda)$. 
In Fig.~\ref{Isca_vs_A_comb} we demonstrate this concept: we have
moved the dark-core data points to their best-fitting 'scattering-equivalent' extinctions 
in the left-hand part of the diagram. A choice of $A^{\rm sca}(y) = 5.0$\,mag and the corresponding 
$A^{\rm sca}(\lambda)$ values for the other bands, resulting from the \citet{cardelli89} $R_V=3.1$ 
extinction law, brings them to a good alignment with the $A(\lambda) < 6$\,mag data points.

A least-squares fit to the 384 nm data points was performed using equation (\ref{Eq1}) and the 
{\tt mpfitfun}\footnote{www.physics.wisc.edu/$\sim$graigm/idl/fitting.html} fitting program. The 
resulting curve is shown as the red line in  Fig.~\ref{Isca_vs_A_comb} and it is seen to be a 
good representation { for the bands 416 nm, $b$ and $y$ as well. The curve for the $u$ band,
on the other hand, deviates slightly from the other curves. 
When adjusted to the 384 nm 'standard curve' at low and intermediate extinctions, 
$A(\lambda)\la 2$\,mag, the $u$ band curve falls below the other curves at $A(\lambda)\ga 4$\,mag.
A possible explanation for this is that, while the dust albedo  $a_{\lambda} \sim const$  over 
$\lambda \sim 380 - 600$ nm, both observations and modelling suggest that 
$a_{\lambda}$ is lower in the $u$ band ($\sim 350$\,nm), see e.g.  \citet{mat18, togi, hensley}. 
Scattered light from high-extinction sight lines,  
$A(\lambda) \ga 4$\,mag, has suffered multiple scatterings (and absorptions). 
At each scattering event the intensity is reduced by a factor of  $a_{\lambda}$. 
The intensity of the scattered light escaping from the cloud is thus proportional to $a^n_{\lambda}$,
where $n = n(\tau)$, the number of scatterings, increases with increasing optical depth $\tau$.  
A smaller albedo in the $u$-band thus causes the scattered light intensity 
at large optical depths to drop below the 'standard curve'. 
For numerical modelling results, see \citet{mat18}, Fig. A.2.}  

The different sections of the curve can be understood as follows:\\
(1) For low optical thickness, $\tau(\lambda) \la 0.5$, 
 $I_{sca}$ increases linearly with the LOS dust column density; the slope is determined
by the product of the ambient starlight radiation field intensity, the albedo and the
extinction coefficient,  
$I_{\rm ISRF}({\lambda})\times a_{\lambda}\times k_{\lambda} $,
all of them being wavelength dependent. The observed slopes are given in Table\,\ref{Tab1}.

(2) When the LOS  $A(\lambda)$ reaches  2 to 3 magnitudes $I_{sca}$ becomes saturated and levels 
off to a maximum intensity value which depends on the albedo and the ambient radiation field but
not on the extinction.

(3) For still higher optical depths,  $A(\lambda) > 3-4$\,mag, $I_{sca}$ decreased slowly.
This is due to the internal extinction and increasing number of multiple scatterings of photons 
off dust particles. Because the dust albedo $a_{\lambda}$ is smaller than 1 ($a_{\lambda} \sim 0.6$) 
photons are increasingly being lost in every additional scattering event.


\begin{table}
 \centering
  \caption{Intensities and colours of scattered light in the low-extinction range. The 5-colour photometry
intensities are in units of \cgs\,, the slopes  correspondingly in \cgs\,mag$^{-1}$; 
the $UBV$ intensities are in  S$_{10}$ units and the slopes in S$_{10}$\,mag$^{-1}$.}
  \begin{tabular}{llrlr}
\hline
 $A_{\rmn V}$& \multicolumn{2}{l}{Slope $I(\lambda)$ vs  $A_V$}& \vline \quad    Colour            &              \\
\hline  
$<0.7$      & $I(u)$    &$22.1\pm3.8$ &\vline \quad   ${I(384)}/{I(u)}$  & $1.38\pm0.26$\\
 $<0.7$     & $I(384)$  &$30.5\pm2.4$ &\vline \quad   ${I(416)}/{I(384)}$& $1.26\pm0.13$\\
 $<0.8$     & $I(416)$  &$35.4\pm2.8$ &\vline \quad   ${I(b)}/{I(416)}$  & $0.83\pm0.08$\\
 $<1.0$     & $I(b)$    &$29.3\pm1.9$ &\vline \quad   ${I(y)}/{I(b)}$    & $0.83\pm0.07$\\
 $<1.2$     & $I(y)$    &$24.4\pm1.4$ &\vline  \quad                     &              \\
\hline
 $<0.7$     & $I(U)$   & $16.2\pm1.2$ &\vline \quad  $U - B$ (mag)       &  $-0.29\pm0.09$\\
 $<0.9$     & $I(B)$   & $12.3\pm0.5$ &\vline \quad  $B - V$ (mag)       &  $0.25\pm0.08$ \\
 $<1.2$     & $I(V)$   & $15.5\pm1.0$ &\vline \quad                      &                \\
\hline
\end{tabular}
\label{Tab1}
\end{table}

\subsection{Colour of scattered light in diffuse, translucent and dense areas}

\begin{figure} 
\vspace{-50pt}
\hspace{-2cm} 
 \vspace{-1.5cm}
\includegraphics[width=125mm, angle = 0]{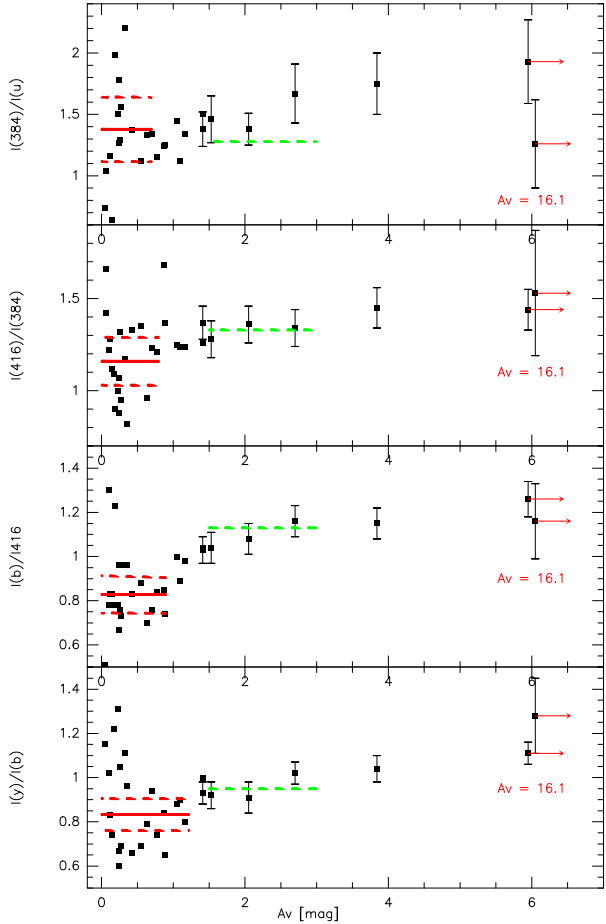}
\caption{Surface brightness ratios in dependence of visual extinction $A_V$ for five optical bands 
from $u$ to $y$ in the LDN\,1642 area. { Error bars are shown for individual positions with  
$A_V \ga 1.5$\,mag. At  $A_V < 1$\,mag the scatter of the values reflects the substantially larger
errors in this range.} The colours from the slopes of $I_{sca}$ vs  $A_V$ relationships 
are shown as red horizontal bars, with error limits as dashed bars. The colours of the mean 
integrated starlight model, as seen by an 'observer' at $z$ = -50 pc, for the Galactic
southern hemisphere ($b < 0$) are also shown by the dashed green lines.  Values for two adjacent
positions in the high extinction core with $A_V = 16.1$ are shown displaced to $A_V = 6$\,mag. }
\vspace{0pt}
\label{Iratios_vsAV}
\end{figure}

The colours of the scattered light were calculated in two ways: 

(1) for all 35 observed positions the intensity ratios $I(384)/I(u), I(416)/I(384), I(b)/I(416)$ and 
$I(y)/I(b)$ were calculated together with their statistical errors; they are plotted in Fig.~\ref{Iratios_vsAV}. 

(2) at low $A_V$ the relative errors of the ratios became disturbingly large; 
therefore, for the sample with  $A_V \la 0.7 - 1$\,mag, representing the 'diffuse dust', 
the colours were derived from the slopes of $I_{sca}$ vs  $A_V$ relationships; they are given in Table 1.
From these slopes we obtain the colours, $I(384)/I(u)$ etc, as given in the Table.  
They are also shown in Fig.~\ref{Iratios_vsAV} as the red horizontal bars, together with their 
error limits as dashed bars. 

The broadband $UBV$ intensities can be composed as weighted sums of the
5-colour intensity values:\newline
$I(U) = 0.471\, I(u)+ 0.231\, I(384) + 0.026\, I(416)$\newline
$I(B) = 0.059\, I(384) + 0.215\, I(416) + 0.185\, I(b) + 0.015\, I_{y}$\newline
$I(V) = 0.097\, I(b) + 0.733 \, I(y) $ \newline
The 5-colour intensities are in \cgs\, and the resulting UBV intensities 
in S$_{10}$ units (see \citealt{mat96}).

\begin{figure} 
\vspace{-40pt}
\hspace{-2cm}
 \vspace{-1.5cm}
\includegraphics[width=125mm, angle = 0]{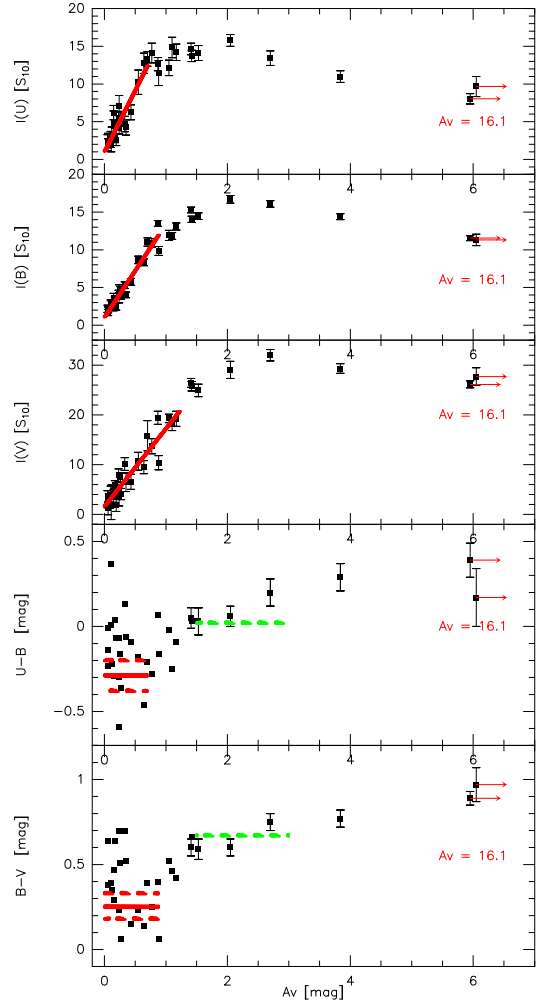}
\caption{UBV surface brightnesses and $U-B$ and $B-V$ colours in dependence of visual extinction 
$A_V$ in the LDN\,1642 area.  The $I_{sca}$ vs  $A_V$ relationships for  $A_V \la 1$\,mag are shown
for U, B, and V; the  $U-B$ and $B-V$ colours derived thereof are shown as the red horizontal bars, with 
error limits as dashed bars. { $U-B$ and $B-V$ error bars are also shown for individual positions with  
$A_V \ga 1.5$\,mag. At  $A_V < 1$\,mag the scatter of the values reflects the substantially larger
errors in this range.}  The $U-B$ and $B-V$ colours of the mean integrated starlight model,
as seen by an 'observer' at $z$ = -50 pc, for the 
Galactic southern hemisphere ($b < 0$) are also shown by the dashed green lines. Note that Vega magnitudes 
are used.  Values for two 
adjacent positions in the high extinction core with $A_V = 16.1$ are shown displaced to $A_V = 6$\,mag. }
\vspace{10pt}
\label{UBV_vsAV}
\end{figure}

The resulting $UBV$ intensities together with the colour indices $U-B$ and $B-V$ are shown in Fig.~\ref{UBV_vsAV}.
In the linear range at small optical depths the slopes were again determined and used to derive
the colour indices $U-B = -0.29\pm0.09$\,mag, and $B - V = 0.25\pm0.08$\,mag, as given in Table 1. 
They are also shown in Fig.~\ref{UBV_vsAV} as the red horizontal bars together with their errors limits  
as dashed bars.

The observed colours from both the five-band as well as the $UBV$ photometry follow, qualitatively, the
same trends: \newline
(1) at small and moderate optical depths, $A_V \la 1$\,mag, representative of cirrus and 
diffuse dust areas the colours are at the blue end of their range; with increasing  $A_V$ they turn redder
up to  $A_V \sim 1.5$\,mag; \newline
(2) in the translucent dust range,   $1.5 \la A_V \la 3$\,mag, the colours reach saturation values which
change only little over this range; this reflects the behaviour of the $I_{\rm sca}$ vs  $A_V$ curves 
 which also reach their saturation (and maximum) level in this range (see Fig.~\ref{Isca_vsAV});\newline
(3) moving towards the dense/opaque dust dust areas,  $A_V \ga 3$\,mag, the colours continue to redden,
though very slowly in most cases.

The background light in the interstellar space outside our solar system is dominated by 
 starlight scattered off interstellar dust along the LOS. Our results in the LDN~1642 area 
demonstrate that, depending on the LOS dust opacity, the background can assume a colour from a 
 wide range of values.

\begin{figure} 
\vspace{-40pt}
\hspace{-1.5cm} 
\vspace{-6.0cm}
\includegraphics[width=125mm, angle = 0]{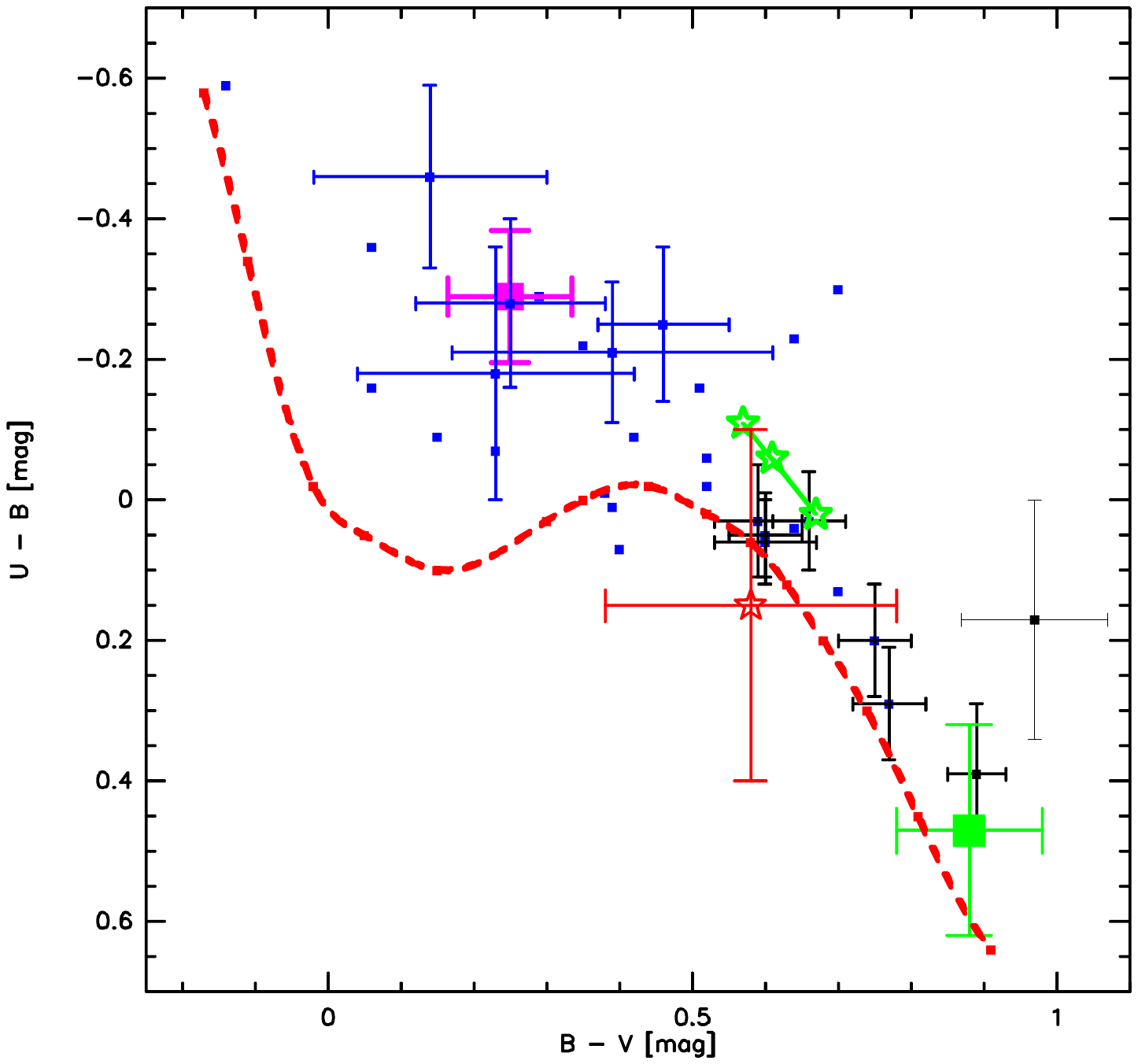}
\caption{Colour-Colour diagram $U-B$ vs $B-V$ of the surface brightness in the LDN\,1642 area. 
Black dots with error bars: $I(V)>20$ S$_{10}$; blue dots with error bars: $15<I(V)<20$ S$_{10}$; 
magenta dot corresponds to the colours derived from the slopes $I$ vs $A_V$. The blue dots without error bars
are for $I(V)<15$ S$_{10}$ where the colours of individual positions cannot be meaningfully derived.
The green asterisks from upper left to lower right indicate the colours derived from the synthetic ISL model 
of \citet{Lehtinen13} for a virtual observer position at $z = -50$\,pc off the Galactic plane
for $b > 0$, all sky, and $b<0$.
The red dashed curve shows the main sequence location of stars, spectral types from B5\,V to K2\,V.
The red asterisk with error bars indicates the colours of the scattered light envelope of the carbon star 
IRC~+10216 \citep{mauron03}. The green square at lower right is for the colours of  the stream 
associated with the galaxy NGC\,5907 (\citealt{laine,dokkum}). }
\vspace{-10pt}
\label{UBvsBV}
\end{figure}

A convenient way to demonstrate this further is by means of the $U-B$ versus $B-V$ Colour-Colour diagram, 
commonly used for illustration of the colours of stars. We have plotted
in Fig.~\ref{UBvsBV} the  $U-B$ vs $B-V$ values of all our observed positions in the LDN~1642 area. 
The black dots with error bars are for positions with $I(V)>20$ S$_{10}$ ($A_V > 1.3$\,mag); 
the blue dots with error bars are for $15<I(V)<20$ S$_{10}$ ($0.7 < A_V < 1.3$\,mag ); 
 the blue dots without error bars are for positions with $I(V)<15$ S$_{10}$ 
($A_V < 0.7$\,mag); these are positions for which the colours of individual positions cannot be accurately 
derived. The magenta square with error bars corresponds to the colours derived from the slopes 
for $A_V < 1.2$\,mag. 

The three green asterisks indicate the colours of integrated starlight derived using 
a synthetic model of the local Galactic environment \citep{mat80a,mat80b,Lehtinen13}; 
the 'observer' is at $z=-50$\,pc, corresponding to the location of LDN~1642; 
the asterisks from upper left to lower right indicate the values for the northern 
Galactic hemisphere ($b > 0$), the whole sky, and the southern hemisphere ($b < 0$).
For comparison, also the location of the main sequence of stars is schematically indicated 
with red dashed curve. The colours of the mean integrated starlight for $b > 0$ are also shown
in Fig.~\ref{UBV_vsAV} by the dashed green lines.  
The red asterisk with error bars indicates the colours of the carbon star envelope IRC~+10216 
(\citealt{mauron03}, and Mauron 2022, private communication).
These comparisons, along with others, will be discussed in section 4.1.

\subsection{Spectral energy distributions of scattered light}

\begin{figure} 
\vspace{-20pt}
\hspace{-0.8cm}
 \vspace{-2.5cm}
\includegraphics[width=80mm, angle = -90]{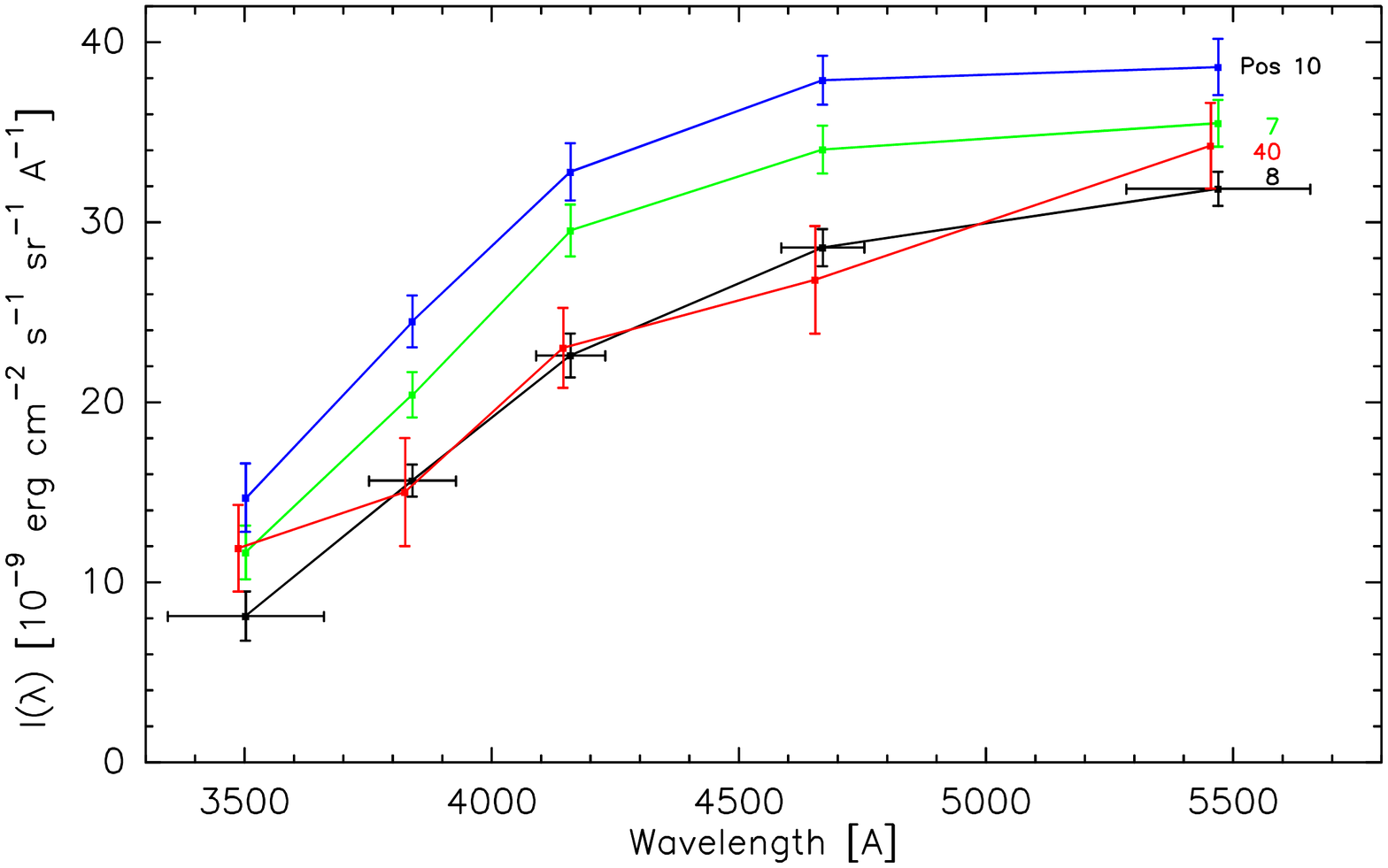}
 \vspace{-2.5cm}
\hspace{-0.8cm}
\includegraphics[width=80mm, angle = -90]{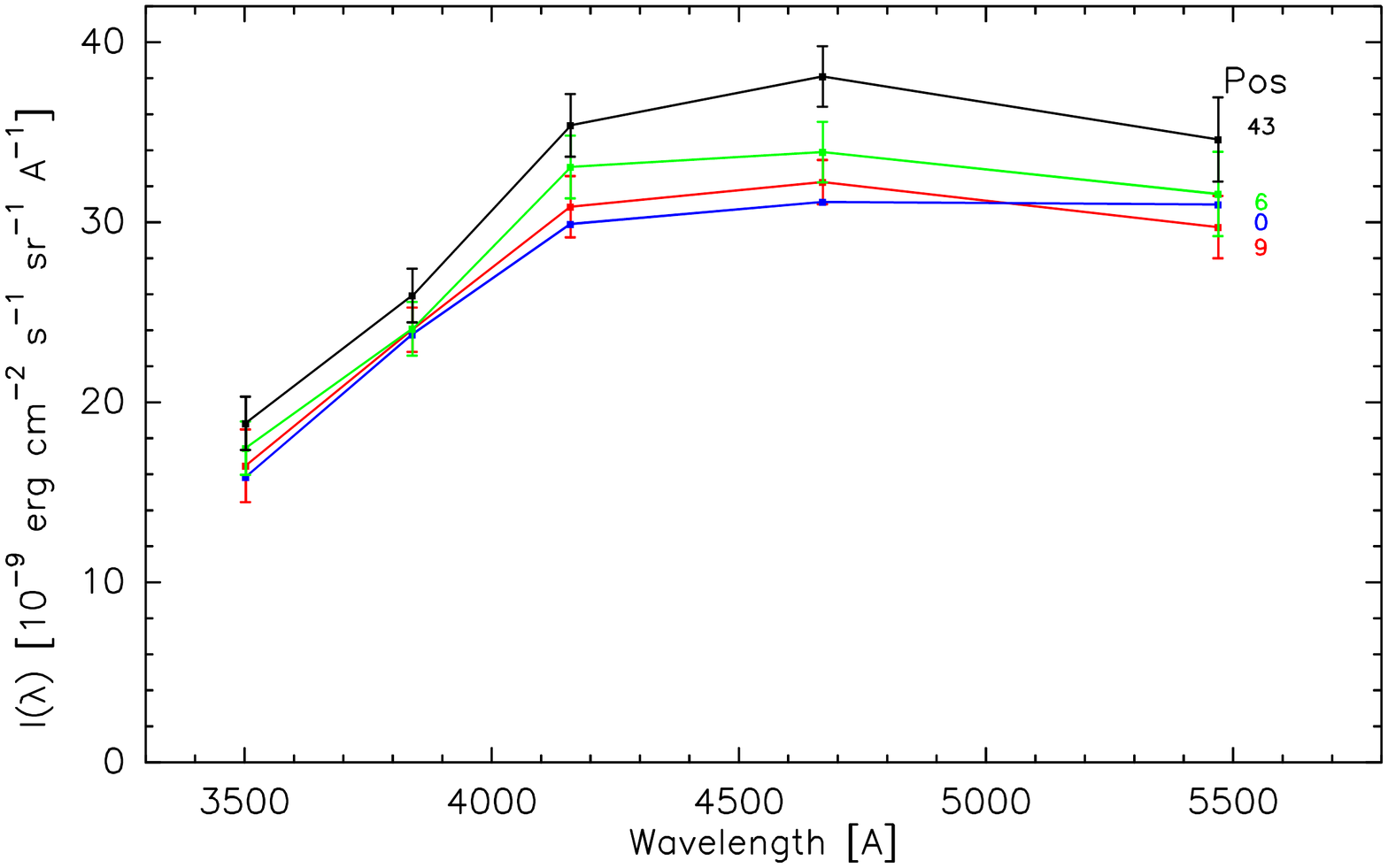}
 \vspace{-2.5cm}
\hspace{-0.8cm}
\includegraphics[width=80mm, angle = -90]{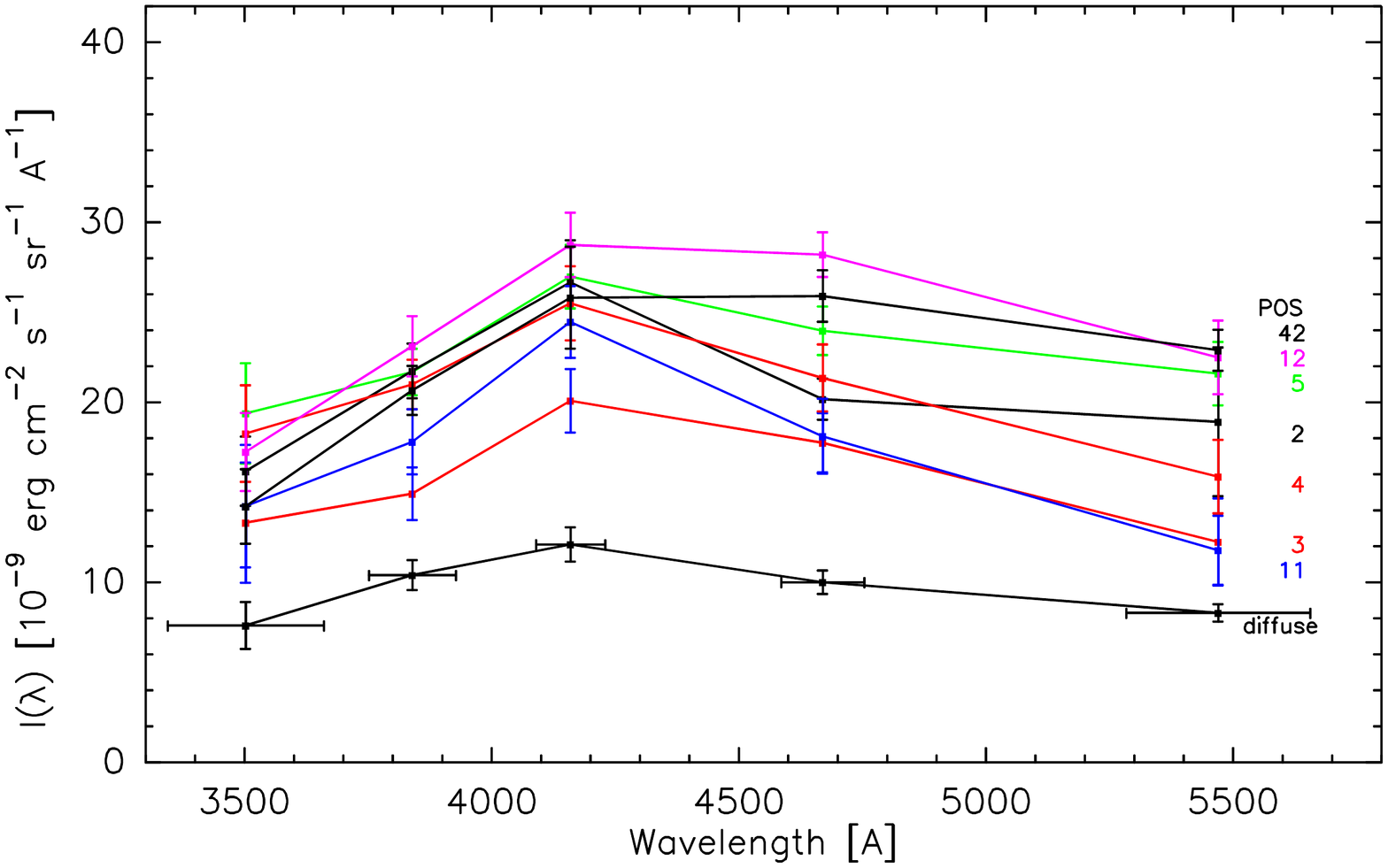}
\vspace{2.0cm}
\hspace{-0.8cm}
\caption{SED's of scattered light for positions in the central and circum-central area of LDN\,1642.
{\em upper part:}  Four high-opacity positions; {\em middle part:} four 
intermediate-opacity positions; {\em bottom part:} seven modest-opacity, $A_V \sim 1$\, mag, positions.
The bottom-most SED has been derived for ``diffuse dust'' from the slopes of  
$I_{sca}(\lambda)$ vs $A_{V}$ diagrams. To compare its shape with the other SED's the curve has been 
normalized to $I_{\lambda}(b)$=20\,\cgs\, and shifted down by 10 units.
HPW's of filter bandpass are shown for Pos\,8 and ``diffuse dust'' data points.}
\label{I_vslam}
\end{figure}

Based on our five-colour photometry we present in Fig.~\ref{I_vslam} spectral
energy distributions, $I_{sca}(\lambda)$ vs $\lambda$, over the wavelength range $\lambda = 350 - 550$\,nm. 
Three groups of SED's are shown; they cover four LOS extinction ranges: \newline 
(1) positions 7, 8 10 and 40 with  
$A_V = 2.7 - 16$\,mag (top panel);\newline 
(2) positions 0, 6, 9 and 43 with $A_V = 1.4 - 2.0$~mag (middle panel); and \newline
(3) positions 2, 3, 4, 5, 11, 12, and 42 with  $A_V = 0.55 - 1.2$~mag (bottom panel). \newline
(4) In addition, the SED for low-opacity diffuse dust is shown in the bottom panel. In this case, 
instead of using individual positions, the SED is based on the slopes of the $I_{sca}(\lambda)$ vs $A_V$ 
relations (see Table 1). 
The values have been scaled to $I_{sca}(b)= 10.0$~\cgs\,.

A qualitative look at the SEDs shows a systematic behaviour: the SED's become increasingly
redder as the LOS extinction increases. 

\begin{figure} 
\vspace{-20pt}
\hspace{-0.8cm}
 \vspace{-2.5cm}
\includegraphics[width=80mm, angle = -90]{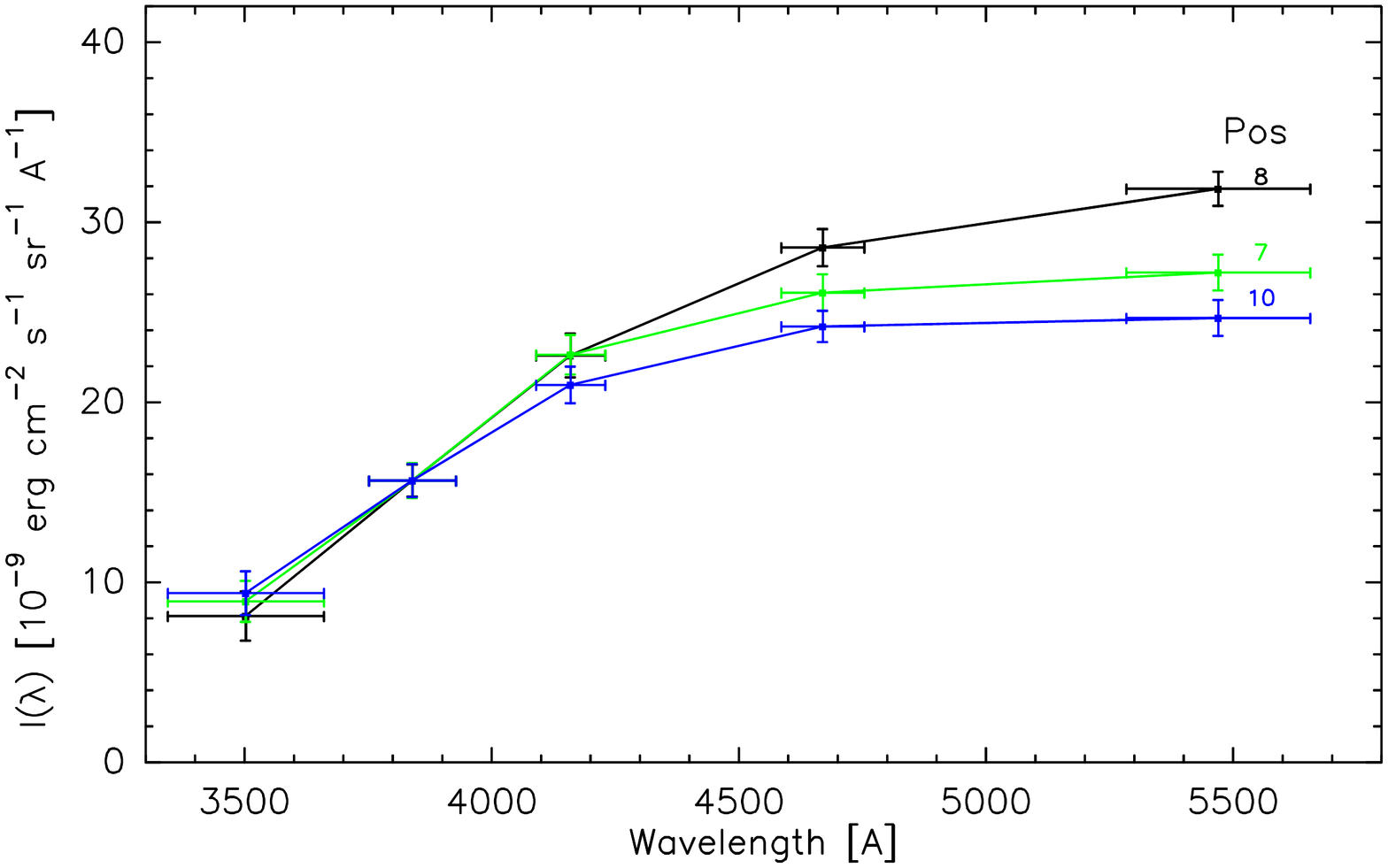}
\vspace{-0.5cm}
\hspace{-0.8cm}
\includegraphics[width=80mm, angle = -90]{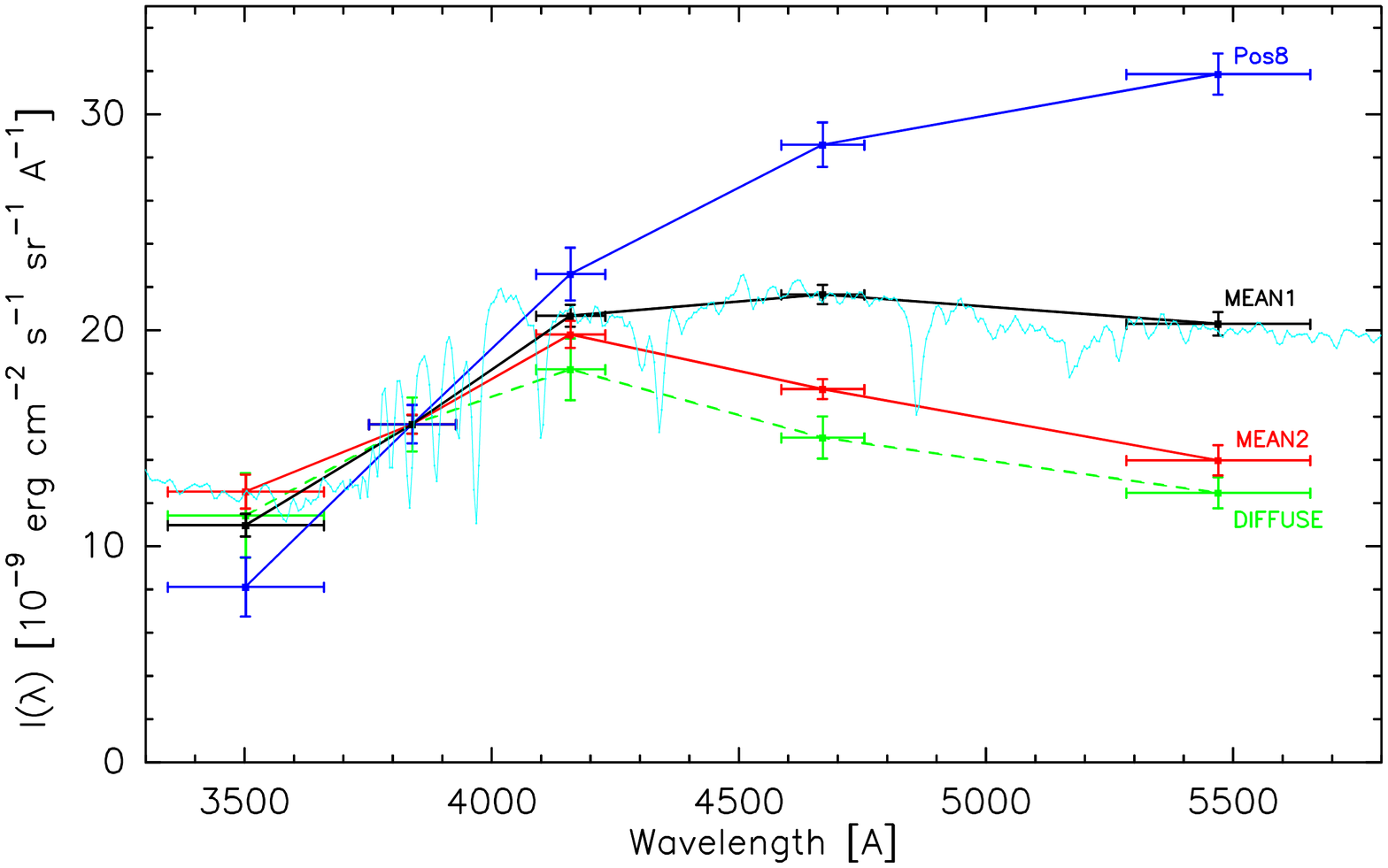}
\vspace{-0.5cm}
\caption{SED's of scattered light normalized to Pos8 at 384 nm.
{ The {\em upper panel} shows the SED's for three high-opacity positions, $A_V = 2.7 - 16$\,mag.
The {\em lower panel} shows averages of the SED's for two groups, 
$A_V = 1.4 - 2.0$~mag and  $A_V = 0.55 - 1.2$~mag which are designated as  MEAN1 and MEAN2. 
Also shown are the statistical errors, calculated from
the errors of the individual SED's included in each mean. 
For comparison, we show also the SED's for Pos 8 and the diffuse dust, as in Fig. \ref{I_vslam}.
The spectrum of the ambient 
integrated starlight radiation field is shown as the light blue line; it is the synthetic 
spectrum, mean over the sky.}}
\label{I_vslam_norm}
\end{figure}

In Fig.~\ref{I_vslam_norm}\, {\em upper panel} we compare the SEDs for the three high-opacity positions
which have been normalized to the value of Pos\,8 at 384 nm. For the lower-opacity positions 
we show in Fig.~\ref{I_vslam_norm}\,{\em lower panel} averages of the SEDs for two groups, 
$A_V = 1.4 - 2.0$~mag and  $A_V = 0.55 - 1.2$~mag, which are designated as  MEAN1 and MEAN2. 
(see Fig.~\ref{I_vslam}). Also shown are the statistical errors, 
calculated from
the errors of the individual SEDs included in each mean. 
For comparison, we show also the SEDs for Pos 8 and the diffuse dust. The spectrum 
of the ambient integrated starlight radiation field is shown as the blue dotted line; it is the synthetic 
spectrum, mean over the sky, as calculated for an observer's location 50 pc off the Galactic plane 
(see \citealt{Lehtinen13}). It is seen to agree closely with the SED of MEAN1.

As demonstrated by Fig.~\ref{I_vslam_norm} the shape of the SED is influenced by the LOS extinction.
For a given position, the cloud's optical thickness decreases towards longer wavelengths. Thus, for the small 
optical depth positions with $A_V \la 1$, the scattered light intensity decreases towards longer 
wavelengths. On the other hand, for large-optical-depth positions with $A_V \ga 4$, $I_{sca}$ becomes
saturated at shorter wavelengths while it still is on the increasing branch at the longer 
wavelengths.

In the presence of a surface brightness contribution from beyond the cloud, such as the EBL, 
another effect will enter that influences the SED.
For a wavelength slot where $I_{sca}(\lambda)$ has a strong discontinuity, such as the 4000~\AA\, break,
the observed SED is expected to differ from the scattered light SED
in a marked way. This effect can be partly the reason for an increase of
the 4000~\AA\,  jump with increasing extinction (see Figs. ~\ref{Iratios_vsAV} and ~\ref{I_vslam_norm}).

\section{DISCUSSION}

\subsection{Intensity and colour of scattered light in LDN~1642}

The  $I_{\rm sca}(\lambda)$ vs $A(\lambda)$ curves as shown in Figs.~\ref{Isca_vsAV},
\ref{Isca_vsAlamda}, \ref{Isca_vs_A_comb} and \ref{UBV_vsAV} are characterized by the three sections: 
linear rise at $ A(\lambda)\la 0.7- 1$ mag, saturation and maximum at 
$A(\lambda) \sim 1.5- 3$ mag, and a slow decrease toward higher extinctions 
at $A(\lambda) \ga 4$ mag. This curve form is 
physically well understood and has been discussed already in \citet{mat70} and \citet{devries85}. 
First part: cloud is optically thin whereby $I_{\rm sca}(\lambda)$ is proportional to 
the LOS optical depth; second part: the cloud is becoming optically thick; third part: the impinging photons
experience multiple absorption/scattering events before escaping from the cloud, causing
decrease of the surface brightness. 

This surface brightness distribution corresponds also to the structure seen in globules
\citep{witt90,fitz}: a dark core coinciding with the opacity maximum, a bright regular rim 
with elliptical or 
spherical shape at a LOS opacity of $\sim$2; and a gradually falling, optically thin outer halo.  
The angular size of the bright rim increases towards longer wavelengths, corresponding to the 
outwards shift of the  $I_{\rm sca}(\lambda)$ maximum  with increasing wavelength, as seen in 
Fig.~\ref{Isca_vsAV}. 

For a quantitative modelling of the $I_{\rm sca}(\lambda)$ vs  $A(\lambda)$ curves the two
basic input ingredients are the impinging Galactic radiation field, and a radiative transfer 
code including multiple scattering. Dust grain models with given scattering properties,
characterized by the albedo $a_{\lambda}$ and the asymmetry parameter $g_{\lambda}$  of the 
scattering function \citep{hg}, can then be tested 
against the observations. Such modelling results, based partly on the same photometric 
data as in this paper, have already been presented for LDN~1642 \citep{lau, mat18}. 
The results indicate that  $a_{\lambda} = 0.50 - 0.60$, depending on the adopted
value of $g_{\lambda}$ in the range 0.6 - 0.8. 

In Table~\ref{compilation} the mean values of the colours $U-B$ and $B-V$ are given for five
extinction slots in the LDN~1642 area, ranging from cirrus ($A_V \le 1$\,mag) up to dense areas
($2 \le A_V \le 4$\,mag). The individual positions are also shown in Fig.~\ref{UBV_vsAV}.
There is a clear trend, seen both in the table as well as in Fig.~\ref{UBV_vsAV}, indicating that
the colours become systematically redder with increasing extinction. 

In Fig.~\ref{UBvsBV}
the colours have been plotted in the conventional  $U-B$ vs $B-V$ Colour-Colour diagram.
The diffuse dust positions (blue dots) have blue colours and are located in the upper left hand section whereas the
translucent and high-opacity positions (black points with error bars) have red colours and are located 
in the lower right hand section of the diagram. 
For reference, also the location of the main sequence for stars is shown as the dashed red curve.\\ 

\subsubsection{Colour of the radiation field and the illumination geometry}

The interstellar radiation field (ISRF) of the Solar neighbourhood 
is a composite of the light of common stars and is dominated by the spectral types A to F.
We use the synthetic spectrum of integrated starlight as presented in \citet{mat80a,mat80b,Lehtinen13}.
Full range of spectral types and a clumpy interstellar dust
distribution were included in the model. The location of the 'observer' was put at $z = -50$\,pc 
corresponding to the location of LDN~1642. The resulting mean colours over the entire sky  
were  $U-B = -0.06 $, $B-V = 0.61$; they are indicated with the middle green asterisk 
in Fig.~\ref{UBvsBV}; the upper and lower asterisks are for the northern and southern hemispheres,
respectively. 

In addition to the synthetic starlight models, the colour of the impinging ISRF has been evaluated 
also by using the $B$ and $R$ band background starlight mapping, carried out by 
the Pioneer 10 and 11 space probes at heliocentric distance of 3 -- 5 A.U. \citep{weinberg74, toller}. 
All-sky maps have been published by \citet{gordon}; see also his web page\footnote{ 
http://www.stsci.edu/\~\,kgordon\~\,\\pioneer\_ipp/Pioneer\_10\_11\_IPP.html}.
We have adopted from this web page the data files `{\tt Reformated data with all stars}' which,
according to the {\tt README} file, ``give the measurements *including* all the stars'', 
that is, the stars brighter than $6\fm5$  subtracted by \citet{toller} have been returned. \\

In order to mimic the illumination geometry for a dust cloud observed in the direction $l_0, b_0$ 
we have determined the average $B_P$ and $R_P$ band background intensities in and around 
the direction  $l_0, b_0$. We used a virtual telescope with an `instrumental beam' given by the 
Henyey-Greenstein scattering function for $g = 0.75$. This beam has a FWHP of $\sim 25\degr$. 
After transforming Pioneer $B_P-R_P$ colour to the standard $UBV$ photometric system, the colour 
of the Milky Way background starlight in the direction of LDN~1642 was found to have the 
value $(B-V)_{\rm MW} = 0.61$, in good agreement with the all-sky colour estimate derived from 
the synthetic model. In Table~\ref{compilation} this value towards LDN~1642, together with 
values for other cloud directions, is given in column (7).    

\subsubsection{Scattered light colour depends on dust column density}
The $U-B, B-V$ colours of the translucent dust with $A_V \approx 2$\,mag are seen to best 
agree with the ISRF colours (Figs. 6 and 7). This is because the  
$I_{\rm sca}(\lambda)$ curves become saturated  at $A_V \approx 2$\,mag and then, 
according to equation (1),  the intensities 
are given by $I_{\rm sca}(\lambda) \approx C_{\rm ISRF}({\lambda})\times a_{\lambda}$. 
For a constant albedo, $a_U = a_B = a_V$,
 the scattered-light colours will agree with the colours of the ISRF.
  
In the range  $A_V \la 1$\,mag, that is for diffuse/cirrus dust, the scattered light 
intensity can, according to equation (1), be expressed as  
$I_{\rm sca}(\lambda) = C_{\rm ISRF}({\lambda})\times a_{\lambda} \times k_{\lambda}\times l$  where
the LOS layer thickness $l$ does not depend on $\lambda$. Assuming a constant albedo,
 the colours of the diffuse/cirrus dust will be given by: \newline
$(U-B)_{\rm Cirrus} = (U-B)_{\rm ISRF} -2.5 \log (k_U/k_B)$ and \newline 
$(B-V)_{\rm Cirrus} = (B-V)_{\rm ISRF} -2.5 \log (k_B/k_V)$. \newline
For the standard extinction law according to \citet{cardelli89} with $R_V$ =3.1,  
$k_U/k_B = 1.18$ and $k_B/k_V = 1.33$, and 
the $U-B$ and $B-V$ colour indices will be shifted by -0.18 and -0.31 mag relative to the ISRF values.
As seen from Fig.~\ref{UBvsBV}, this amount of bluening is to a good approximation indeed the case
for the observed colours of the diffuse/cirrus dust.

The positions with large extinctions,  $A_V > 3$\,mag, as displayed in Figs. ~\ref{UBV_vsAV} and ~\ref{UBvsBV}, 
show reddening relative to the ISRF colours, both in $U-B$ and $B-V$. { Similarly as} the bluening in
the cirrus areas this reddening can qualitatively be explained in terms of the dust extinction law: While 
the (normalized) $I_{\rm sca}(\lambda)$ vs  $A(\lambda)$ curves as shown in Figs.~\ref{Isca_vsAlamda} 
and \ref{Isca_vs_A_comb} are almost indistinguishable, there are distinct differences when using 
the wavelength-independent optical depth indicator $A_V$ as the x-axis, see Figs.~\ref{Isca_vsAV},
\ref{UBV_vsAV}. According to Fig.~\ref{UBV_vsAV}  the $I_{\rm sca}(V)$ vs  $A_V$ curve has its 
maximum at $A_V \sim 3.0$\,mag. Correspondingly, the maximum in the B band occurs at  $A_B \sim 3.0$\,mag,
which according to the standard extinction law \citep{cardelli89} corresponds to $A_V \sim 2.25$\,mag.
This is also confirmed by  Fig.~\ref{UBV_vsAV}. Beyond this point the  $I_{\rm sca}(B)$ curve
starts to drop whereas the  $I_{\rm sca}(V)$ still continues to rise up to  $A_V \sim 3.0$\,mag;
and, all the way beyond this maximum point its decline lags behind the more rapidly dropping  
$I_{\rm sca}(B)$ curve. This obviously leads to the increase of the $B-V$ colours, and the same is 
happening with $U-B$. Unlike for the bluening in the case of cirrus dust, the amount of the reddening 
cannot be readily derived from the extinction law alone.

\begin{table*}
\begin{minipage}{170mm}
\centering
\caption{U - B and B - V colours for scattered light of high-latitude cirrus and translucent clouds. 
In addition some lines of sight near Galactic plane are included for comparison. }
  \begin{tabular}{lccccccl}
\hline
 Cloud  & $l$    & $b$   & $A_V$      & $U - B$       & $B - V$        & $(B-V)_{\small \rm MW}$
\footnote{Milky background starlight colour from Pioneer 10/11 $B_P, R_P$ maps, averaged with $g=0.75$
 'dust scattering beam' or over a {$\O25\degr$} area} & Ref.\\ 
(1)&(2)&(3)&(4)&(5)&(6)&(7)&(8)\\

\hline  
\multicolumn{3}{l}{\em Dark and translucent clouds} & &&&&\\
LDN~1642& 211.0  &-36.5  & 0.0-1.0    & $-0.29\pm0.09$   & $0.25\pm 0.08$ & 0.61      &1   \\
        &        &       & 0.55-0.88  & $-0.21\pm0.03$   & $0.23\pm 0.07$ & 0.61      &1   \\
        &        &       & 1.05-1.17  & $-0.12\pm0.07$   & $0.47\pm 0.03$ & 0.61      &1  \\
        &        &       & 1.41-1.53  &\,\, $0.04\pm0.05$& $0.62\pm 0.03$ & 0.61      &1  \\
        &        &       & 2.05-3.84  &\,\, $0.22\pm0.04$& $0.83\pm 0.07$ & 0.61      &1   \\

Draco Nebula & 90.0   & 39.0  &  $\le 1.1$ &\, \, $0.00\pm0.03$$^b$& \, $0.69\pm 0.09$$^b$ & 0.79 & 2  \\
LDN~1780     & 359.0  & 37.5  & $\le 0.9$&             & \, $0.86\pm 0.03$\footnote{using  $I_{\rm obs}$ vs  $A_V$ slopes}&0.89 & 2 \\
MBM\,25   & 173.8     & 31.5  & 0.9  &  & $0.66\pm 0.08$ &  0.72    & 5   \\
MBM\,30   & 142.2     & 38.2  & 0.5  &  & $0.97\pm 0.12$ &  0.75    & 5   \\
MBM\,32   & 147.2     & 40.7  & 0.6  &  & $0.63\pm 0.14$ &  0.75    & 5   \\
MBM\,41A (=Draco Nebula)   & 90.0     & 39.0  & 0.6  &  & $0.76\pm 0.12$ & 0.79     & 5   \\
MBM\,41D   & 92.3    & 37.5   & 0.5  &  & $0.77\pm 0.15$         & 0.79     & 5   \\

Ring Nebula&183.3& 25.1  & $\le 0.27$ &$0.39$         & $0.51$         &  0.71    & 6 \\
Virgo/NGC4435 plume  & 280.2         & 74.9         & $\la 0.3$    &   & 0.5 - 0.75&  0.78&  3\\
Thumbprint bright rim   &302.6   & -15.9 & $\sim2$\footnote{estimated from $A_V$ map of \citet{kainu}} &  $0.0\pm 0.20$& $0.85\pm 0.18$ & 0.78     & 4   \\
\hline
\multicolumn{3}{l}{\em Widely distributed cirrus} & &&&&\\

SDSS cirrus&  $0...360$   & $|b|=35...50$& $\la 0.5$&      & $\sim 0.62$\footnote{estimated from Fig.~4 in \citet{brandt} }  &0.86 & 9 \\
           & $|l|<60\degr$ &-90...+90      &  $\la 0.5$&     & $\sim 0.71$$^e$& 0.94 & 9 \\
SDSS Stripe82 &  $55...101$ & $-61...-33$& 0.2 - 0.4 &   & $0.65\pm0.02$\footnote{using \citet{cook} colour transformations 
from SDSS $ugri$ to Johnson-Cousins $UBVR_c$ photometry for nearby galaxies}  &0.86\footnote{for $l=65\degr, b=-40\degr$} & 10  \\
              &  $186...190$& $-41...-38$& 0.4 - 1.3 &   & $0.81\pm0.05$$^f$ &0.68\footnote{for $l=188\degr, b=-40\degr$} & 10  \\
                 
\hline
\multicolumn{3}{l}{\em Dust envelopes of AGB stars} & &&&&\\
IRC+10216 &  221.4   & 45.1&   $\sim2$ & $0.15\pm0.25$&$0.58\pm 0.20$ & 0.72   &11 \\
IRAS16029-3041 & 345 & 16  &   $\sim2$ &              &1.00\footnote{corrected for foreground reddening using http://ned.ipac.caltec.edu} & 0.94  &11 \\
IRAS17319-6234 & 330 & -16 &   $\sim2$ &              &1.06$^h$ & 0.89  & 11\\
\hline

\end{tabular}
\label{compilation}

{\bf References:} {(1) this paper; (2)~\citet{mat18}; (3) \citet{rudick}};
(4) \citet{fitz}; \\(5) \citet{witt08}; (6) \citet{zhang};
(9) \citet{brandt}; (10) \citet{roman};\\(11) \citet{mauron03,mauron13}; Mauron(2022, private communication)
\vspace{-0.2cm}
\end{minipage}
\end{table*}

\subsection{Comparison with a cirrus/translucent cloud sample}

To compare with LDN~1642, we have collected in Table 2 colour data for a sample of
diffuse and translucent clouds. All but one of these clouds are at high galactic latitudes,
$|b|\ge 25\degr$. Besides $B-V$, also the $U-B$ colours are available in a few cases. 

For  Draco Nebula and LDN~1780, the colours were
determined using slopes fitted to $I_{\rm sca}$ vs  $A_V$ curves in the 
diffuse dust domain,  $A_V \la 1$\,mag. For the sample of five MBM catalogue clouds the
 $B-V$ values are based on the mean surface brightnesses in two bands, $B$ and $G$, averaged over several 
positions in each cloud \citep{witt08}. The corresponding mean extinction, 
column (4), puts them into the optically thin dust domain, as well. The TPN bright rim represents 
a case with a clearly larger optical depth. From deep $HK$ band photometry \citet{kainu} 
found that the bright rim follows approximately the $A_V \approx 2 -3$\,mag extinction contours.
The TPN rim can, therefore, be compared with the translucent dust areas in LDN~1642.

We have included in Table 2 two
wide-field studies of high-galactic-latitude diffuse/cirrus clouds \citep{brandt, trujillo}.
They are both based on archival Sloan Digital Sky Survey (SDSS) data.

The colour of the scattered light from a dust cloud depends on the colour
of the ISRF light impinging on it. Since all the clouds, except perhaps for 
the Draco Nebula (= MBM 41D), are nearby, they are embedded in the Solar neighbourhood ISRF. 
Because of the forward-throwing 
scattering function of the dust grains their {\em received illumination} depends on the 
direction $l_0, b_0$ in which the cloud is seen. Using  the Pioneer 10/11 data base
we have calculated for each cloud the colour of the 
background starlight, $(B-V)_{\rm MW}$, which is taken to be the average over a 25\degr diameter 
area of the Milky Way, centered on $l_0, b_0$, the direction of the cloud. The values are given in column (7).
For the two wide-field cirrus surveys, mean values 
corresponding to their sky coverage have been estimated. 

Are there any general trends to be seen in the colours? One expects that 
the clouds become redder towards the Galactic centre where the contribution by the red central 
bulge increases \citep{toller}. To test this trend there is one cloud, LDN~1780, 
that is located at $l = 359\degr$. It is, indeed, substantially redder ($B-V = 0.86$)
 than its anti-centre counterparts LDN~1642 ($B-V = 0.3 - 0.5$),  MBM~25 ($B-V = 0.66$)
and Ring Nebula ($B-V = 0.51$).
Also the SDSS cirrus data of \citet{brandt} show this effect: $B-V \approx 0.71$ towards
the central section ($|l| < 60\degr$) as compared to $B-V \approx 0.62 $ for the stripe
at $|b| = 35\degr - 50\degr$ over all longitudes.

The colours of the 'SDSS cirrus' and the 'SDSS Stripe\,82' appear to behave
similarly as the optically thin dust in LDN\,1642, that is, $(B-V)$ is 
smaller than  $(B-V)_{\rm MW}$.

\subsection{Comparison with circumstellar envelopes of AGB-stars}

 High mass-loss rates of some asymptotic giant branch (AGB) stars have created around them 
extended envelopes of gas and dust that completely obscure the central star in the optical.
Such envelopes are externally illuminated  by the Galactic radiation field. 
The best known object in this class is IRC +10216 \citep{mauron03};
 currently, some 20 such envelopes are known (\citealt{mauron06,mauron13}).    

We have listed in Table\,\ref{compilation} three AGB envelopes for which the $B-V$ colours 
have been determined. IRC\,+10216, for which also the $U-B$ colour is known 
\citep{mauron03},  is marked with a red asterisk in 
Fig.~\ref{UBvsBV}; its colours agree with the local ISRF model and
with LDN~1642 in the range $1< A_V <2$\,mag. Like LDN\,1642 and the other clouds in Table 2, 
IRC\,+10216, with its distance of 120 pc, is exposed to the local ISRF. 

The two other AGB stars, IRAS16029-3041 and IRAS17319-6234, are more distant, at 1900 pc and 1200 pc,
respectively. 
Being located towards the Galactic bulge direction these two objects are exposed, 
like LDN~1780, to a radiation field redder than the other objects in the sample.
Their red colours,  $B-V$ = 1.0 and 1.06\,mag,  may partly reflect  
this radiation field property. Their colours are redder than the Milky Way background starlight colours 
of $(B-V)_{\rm MW}$ = 0.94 and 0.89,  which is as expected for an optical thickness of $\sim 2$. 

The difference in colour between the IRAS16029-3041 and IRAS17319-6234 envelopes on one side, and IRC\,+10216
on the other, may also reflect the difference of their dust compositions. The first
two have silicate dust while for IRC\,+10216 the dust is carbon rich.

The illumination geometry of IRAS16029-3041 and IRAS17319-6234 differs from the other objects in our sample:
because of their large $z$-distances of 520 and -330 pc 
they are located outside of the Galactic layer of young stars and dust. 
From their vantage points they  'see' a wider Galactic area than the local clouds do. 
Because of the forward-throwing 
scattering function of dust 
they scatter toward us mostly light from the outer Galactic disk and halo,
substantially redder than the inner disk. 
Another, intriguing factor is that these two AGB stars 'see' the bright and red light from the 
Galactic central area, un-obscured by the massive intervening extinction that prevents our view. 
They scatter light from the Galactic centre 
 toward us at the favorably small scattering angles 
of $\sim 25\degr$ and $\sim 35\degr$, respectively.  This view is supported also by 
the finding of  \citet{mauron13} that these two envelopes have the highest  $B$ band surface 
brightnesses in their whole sample, namely  23.8 and 23.9\,mag/$\sq\arcsec$, respectively.

Another reason for the higher surface brightness of the IRAS16029-3041 and IRAS17319-6234
envelopes may well be their location closer to the Galactic centre, at $R_{\rm GAL} \sim$6.2 and $\sim$7\,kpc,
where the ISRF is expected to be enhanced relative to the local ISRF at $R_{\rm GAL} = 8$\,kpc
(see \citealt{mauron13} section 4.1.2 and Fig.\,9).  

\subsection{Extragalactic debris {\em or} Galactic foreground cirrus: can optical colours decide?}

 Occasionally 
 a low-surface-brightness plume appears to be associated with a galaxy or is seen towards 
'empty' space in a cluster or a group of galaxies. Then, the question arises whether it is an accumulation 
of unresolved stars stripped off a galaxy 
or just a randomly located foreground cirrus cloud in the Solar neighbourhood.  

A classical example is Arp's Loop associated with M\,81 \citep{arp}. 
Recent 
far infrared observations with {\em Spitzer}/MIPS \citep{sollima} and {\em Herschel} \citep{davies} 
strongly suggest an interpretation as part of the Galactic cirrus structures 
that are prominently present in the M\,81/M\,82 area \citep{barker, sun}.
Not all regions in sky are covered by high-resolution far-IR observations. In such cases 
optical colours of the plume may offer a viable alternative (see, e.g. \citealt{roman}).

Morphologically similar loops, apparently associated with the nearby galaxies, have been detected
 \citep{martinez22}. The nature of the stellar stream around NGC\,5907 \citep{shang} is well established 
as debris of a tidally disrupted globular cluster or a minor galaxy \citep{laine}. For the brightest 
eastern section of the stream the colour is $g-r = 0.64 - 0.7$\,mag (\citealt{laine,dokkum}) 
corresponding to $B-V= 0.88\pm0.10$ \citep{jester}.
 A 'simple stellar populations' model predicts the colours $U-B = 0.47$ and  $B-V=0.86$ \citep{laine}. 
The position in the $U-B$ vs $B-V$ diagram in Fig.~\ref{UBvsBV}\, is shown as a green square.

As discussed in section 4.1.2 above and demonstrated also by Fig.~\ref{UBvsBV}, 
the colour of a local Galactic dust cloud can be predicted in two steps: 
(1) get the colour of the Milky Way background starlight in the direction of the cloud, 
$(U-B)_{\rm MW}$ and $(B-V)_{\rm MW}$; (2) estimate the optical depth of the cloud; 
assuming that it is a cirrus cloud, $A_V \la 1$\,mag, the colours are predicted to be 
 by $\sim 0.20$ and $\sim0.3$ mag bluer in $(U-B)$ and $(B-V)$, respectively; this bluening 
makes the cirrus colours sufficiently different from the expected colours of galactic 
loops or intra-cluster patches which mostly consist of old stellar population.
More seldom, a foreground cloud would fall into the optically thick domain,  $A_V \ga 1$\,mag.
In such a case its colours would be similar or even redder than $(U-B)_{\rm MW}$ and $(B-V)_{\rm MW}$;
then the colour test would not help to discern it from an extragalactic light patch consisting of 
old stellar population stars.

This 'litmus test' does not take into account that the dust scattering properties may differ 
 from one cirrus cloud to another in the local Galaxy. In view of the relatively constant and 
predictable colours as compiled in Table 2, this does not appear a strong drawback, however.

\section{Summary and conclusions}

For an observer 
beyond the zodiacal dust cloud  the sky background light is dominated by the diffuse Galactic light (DGL),
that is the scattered light off interstellar dust particles 
illuminated by the combined radiation of Galactic stars (integrated starlight, ISL) plus contributions by 
ionized gas. 
The DGL covers the entire sky with a wide range of different environments, from optically thin cirrus dust
over translucent to completely opaque dark clouds. 

In the present paper we have studied the intensity and colour of the scattered light from dust in a 
representative high-galactic-latitude area towards a low-extinction direction in the Galactic 
anti-centre. The area 
contains the opaque cloud core of LDN\,1642 ($A_V \approx 16$ mag) and its translucent
periphery ($A_V \approx 1 - 5$ mag), and an area around it with typical 
diffuse cirrus dust with $A_V \la 1$\, mag. 
Surface brightness was measured at 35 positions in five intermediate band filters covering the
wavelength range 3500 - 5500 \AA\,. Because of the large size, $4.5\degr \times 4\degr$, of the area covered, 
methods were developed
for the elimination of airglow and zodiacal light gradients. Time variations of airglow were corrected
for by using simultaneous measurements with a monitor telescope in parallel with the main telescope (Appendix A).
The extinction was determined by setting up the scaling with 2MASS $JHK_s$ 
colour excesses, and by using dedicated 200 $\mu$m absolute photometry measurements for  improving its precision
and for setting the extinction zero point. In the cloud core deep dedicated $HK_s$ photometry
was used (Appendix B).

The basic observational results are the following: 
The intensity of the scattered light, $I_{\rm sca}(\lambda)$,  depends on dust column density in a characteristic
way: For optically thin dust the intensity first increases linearly for  $A_V\la 1$\,mag, then turns to
a saturation value at $A_V\approx 1.5-3$\,mag; at still larger extinctions,  $A_V\ga 3-4$\,mag, the intensity
turns down to a slow, approximately linear decrease. The $A_V$ value of the intensity maximum shifts
in a systematic way from   $A_V\approx 1.5$\,mag at 3500 \AA\, to  $\sim 3$\,mag at 5500 \AA\, (see Fig.~3).

Instead of using $A_V$ as the common column density measure for all five wave bands, we can alternatively 
use for
each wave band its 'own' extinction, $A(\lambda)$. In this case all  five $I_{\rm sca}(\lambda)$ vs  
$A(\lambda)$ curves are closely similar (see Figs.~4); their maxima now occur at closely the
same same wavelength-specific extinction, $A(\lambda)\approx 3$\,mag; after multiplication 
by a suitable scaling factor all curves are brought to a close agreement with the others (see Fig.~5).    

The  $I_{\rm sca}(\lambda)$ vs $A(\lambda)$ curves offer a straight-forward explanation for
the behaviour of the scattered-light colours as function of $A_V$.
The colour of the scattered light depends, naturally, in the first place on the ISRF colour.
For a wavelength-independent albedo the colour at the peak of the $I_{\rm sca}(\lambda)$ curve
agrees with the  ISRF colour. At the low-column-density side of it the scattered light is
bluer and at the high-column density side redder than the ISRF (Figs. 6-7). These colour changes
are a direct consequence of the wavelength dependence of the extinction.
Fig. 8 illustrates the extinction dependence of the scattered light colour by means of the 
conventional $U-B$ vs $B-V$ Colour-Colour diagram.

We have compared the colours of scattered light in the LDN\,1642 area with the $B-V$ and $U-B$ colours obtained 
in a number of other relevant observational studies: a sample of high-latitude diffuse/translucent 
clouds; two wide-field studies of high-latitude cirrus dust; and three AGB star envelopes 
externally illuminated by the ISL radiation field (see Table 2). 

Besides being an interesting observational resource for interstellar dust studies,
the cirrus/DGL is an unwanted foreground contaminant for extragalactic low-surface-brightness sources. 
Its elimination is crucial especially for objects like outskirts of galaxies, 
intra-cluster light, or the extragalactic background light.
Our results for the colours of diffuse/cirrus dust suggest that optical colours can be useful to
distinguish cases where a diffuse plume or stream, apparently associated with 
a galaxy or a group or cluster, is more likely a local cirrus structure.

\section*{Data availability statement}
The data underlying this article are available in the article and its online summplementary material
as specified in the Supporting information section.

\section*{Acknowledgements}
We dedicate this paper to the memory of of Gerhard von Appen-Schnur (formerly Schnur, 
9.3.1940 -- 13.2.2013). He initiated with KM this project and was the key expert of
the observing sessions at La Silla in all runs.
In 1980 D. Schallwich and in 1987 and 1988 KM was the other observer. 
KL was the observer for the NTT/SOFI programme.
We thank the ESO staff at La Silla for their excellent service. 
We thank N. Mauron for his comments and clarifications concerning the envelopes of
the AGB stars. { We thank an anonymous referee for several useful improvements.} 
KM and KL acknowledge the 
support from the Research Council for Natural Sciences and Engineering (Finland); PV  
acknowledges the support from the National Research Foundation of South Africa and
MH the support from the Deutsche Forschungsgemeinschaft grant DFG CH71/33-1.
This research has made use of the USNOFS Image and Catalogue~archive
   operated by the United States Naval Observatory, Flagstaff Station
   (http://www.nofs.navy.mil/data/fchpix/).

\section*{Supporting information}

Supplementary data are available at $MNRAS$ online.

The coordinates of the observed 35 positions are given in table 
{\sc L1642\_AV+5colour+titleNEW.txt}.\\
The differential surface brightnesses relative to the standard position Po8 
in the five optical filter bands, plus the absolute surface 
brightness at FIR 200 $\mu$m, are given in table 
{\sc Final2020JUL9N.txt}

\appendix

\section{Surface photometry in LDN\,1642 area: methods and reductions}

The data reductions and analysis as presented in this appendix are largely based on 
the MSc thesis  of \cite{Vai94}.

Surface photometry of extended sources differs in several aspects from the stellar photometry:
(1) in addition to the atmospheric line-of-sight extinction correction, 
specific methods are required for eliminination of the atmospheric surface brightness contaminants, including 
the airglow (AGL) and the atmospheric scattered light; (2) the AGL shows substantial time 
variations which are crucial when ON and OFF source measuments are not simultaneous; (3) instumental
straylight from off-aperture stars. 

The photometry was carried out in five filters, 
centred at 360~nm $(u)$, 384~nm, 416~nm, 470~nm $(b)$ and 555~nm $(y)$, using the ESO 1-m and 
50-cm telescopes at La Silla. The 50-cm telescope monitored the airglow variations. 
The observations were made differentially, relative to a standard position (Pos\,8) 
in the centre of the cloud. Subsequently, the zero level was set by fitting in RA, Dec 
coordinates a plane through the darkest positions well outside the bright cloud area.The observations 
were carried out in three sessions: 1980 Dec 06-10, 1987 Dec 15-20, and 1988 Dec 02-11.
The coordinates of the observed 35 positions are given in table 
{\small{\sc L1642\_AV+5colour+titleNEW.txt}} (see section Supporting information). 
and are shown in Fig.\,1.

\subsection{Elimination of airglow time variations: two-telescope technique}

In order to eliminate the influence of the airglow time variations we used a two-telescope technique. The 50-cm 
telescope was used as a monitor and was pointed towards a fixed position 
in the dark nebula area during the whole measuring cycle. The large diaphragm size of $\diameter 7'$ gave 
a large signal as compared to the 88\arcsec diaphragm of the 1-m photometer. The integration times  
were synchronized with those of the 1-m telescope within $\sim$1 second. The 1-m telescope was used 
to measure several individual positions within or outside the dark cloud boundaries 
in an area of $\sim$ 4.5\degr -- 4\degr in size. 
One position in the centre of the cloud was chosen as a standard position.

For each observation, be it standard or program position, the five-colour phototometry 
was run according to an automatic sequence of integrations and filter changes.  
The ratio of the signals from the fixed 50-cm position and the 1-m telescope standard position was formed:
\begin{equation}
R(t) = \frac{I_{\rm 50cm}(t)}{I_{\rm 1m}(t)} 
\end{equation} 
$R(t)$ remained very stable during a whole night despite the temporal variations of the airglow.
 The typical variations in sky brightness over one night were ~ 10 -- 30 per cent. The ratio  $R(t)$ varied typically by no
 more than $\sim$2 per cent. The ratio  was fitted with a Chebysev polynomial of low order; normally an order of up to 3 
was sufficient to give a good fit to observations. After removal of the fitted polynomial the residuals 
were of the order of 1 per cent. The sligth variation of $R(t)$ can be understood because: the zenith distance 
difference between the 50-cm and 1-m positions (being {$\sim15\arcmin$} apart) changed in a systematic way 
over the night; in addition, spatial airglow variations, or 'airglow clouds', can cause small sporadic changes.

Between the standard position observations, normally one to two program positions were observed. 
The goal of the differential surface phototometry was to determine the differences of the program positions, PosN, 
relative to the standard position Pos8:
\begin{equation}
 \Delta I_{\rm 1m}^{\rm Pos N} =  I_{\rm 1m}^{\rm Pos N} -  I_{\rm 1m}^{\rm Pos8}
\end{equation}
Because the standard position observations are, necessarily, at a different time, t, from the program positions we use  
the polynomial fit to the ${I_{\rm 50cm}(t)}/{I_{\rm 1m}(t)}$ ratio, $R_{\rm fit}(t)$, to obtain an interpolated
value of $I_{\rm 1m}^{\rm PosStd}$ at time t: 
\begin{equation}
 \Delta I_{\rm 1m}^{\rm Pos N}(t) = I_{\rm 1m}^{\rm Pos N}(t) - R_{\rm fit}(t) I_{\rm 50cm}(t)  
\end{equation}
Here, as explained above, the 50-cm monitor telescope signal was taken, for each PosN, at exactly the same time as the 1-m 
observation. 

\subsection{Correction for the airmass dependence of airglow and tropospheric scattered light}

In differential surface photometry of an extended sky area the atmospheric and  interplanetary components 
are eliminated as far as they are constant over the  
target area. We need to consider only the differential effects caused by the spatial differences 
over the studied area of $\la$4\degr.
   
The total diffuse sky brightness for Position N observed at airmass $X$ is given by
\begin{equation}
I^{X}_{\rm obs}{\rm (PosN)} = I_{\rm DGL}e^{-\tau(\lambda) X} + I_{\rm ZL}{\rm (PosN)} e^{-\tau_R(\lambda) X} + I^{X}_{\rm ADL} 
\end{equation}
where $I_{\rm DGL}$ and  $I_{\rm ZL}$ are the Diffuse Galactic and the Zodiacal Light intensity, respectively, 
outside the atmosphere; the atmospheric total and Rayleigh extinction coeffiecient is denoted 
by $\tau(\lambda)$ and $\tau_ R(\lambda)$. The atmospheric diffuse light (ADL) intensity, $I^{X}_{\rm ADL}$, 
is the sum of the airglow (AGL) and the tropospheric scattered light, $I^{X}_{\rm SCA}$,  that is the sum of the 
three components, with AGL, ZL and ISL as the sources of illumination for the scattered light
\begin{equation}
I^{X}_{\rm ADL} = I^{X}_{\rm AGL} + I^{X}_{\rm SCA} 
\end{equation}

Because the observations of the standard position, Pos\,8, and a program position, PosN, 
are taken at slightly different zenith distances, a systematic effect arises due to the zenith distance dependence 
of the ADL. Near the zenith the atmosphere can be approximated as plane parallel, 
and at airmasses  $1.0 \le X \la 1.4$ the ADL intensity can be approximated by the linear relationship
\begin{equation}
I^{\rm X}_{\rm ADL} = I^{1.0}_{\rm ADL}[1 + k(\lambda)(X-1)]
\end{equation}
where $I^{1.0}_{\rm ADL}$ is the ADL intensity in zenith ($X = 1.0$). 

The ADL difference between airmasses  $X + \Delta X$ and  $X$ is thus given by 
\begin{equation}
\Delta I^{X+\Delta X, X}_{\rm{ADL}}  = {\scriptstyle\frac{k(\lambda)}{1+k(\lambda)(X-1)}} \Delta X\cdot I^X_{\rm {ADL}} 
   =  C(\lambda)\cdot \Delta X\cdot I^X_{\rm ADL} 
\end{equation}
where $C(\lambda)= k(\lambda)[1+k(\lambda)(X-1)]^{-1}$.
We have estimated the values of $k(\lambda)$ empirically by observing 
a pair of positions from a small to a large airmass difference, $\Delta X$ up to $\sim$0.4.

We obtained the values  $k(\lambda)$ = 0.19, 0.24, 0.34, 0.50 and 0.46   
at $\lambda$ = 350(u), 384, 416, 480(b) and 550~nm (y), respectively.

Using  equation (A4) and recognizing that at high galactic latitudes $I_{\rm DGL}$ is much smaller that$I_{\rm ZL}$ 
we obtain  for the ADL correction, referred to outside the atmosphere, the expression
\begin{equation}
\Delta I_{\rm ADL}^{\rm corr} \approx [I^X_{\rm obs} -I_{\rm ZL}e^{-\tau_R(\lambda) X}]C(\lambda) \Delta X e^{\tau(\lambda) X}
\end{equation}
For typical airglow differences, $\Delta X \la 0.02$, the corrections for the five filter
bands ranged up to $\Delta I_{\rm ADL}^{\rm corr} \la 1.6, 2, 2, 1.5$, and $1.3$ \cgs\, 
in $u, 384, 416, b$ and $y$, respectively.

\subsection{Instrumental straylight from off-axis stars}

The surface brightness observed towards blank areas of sky, even if
the photometer aperture is devoid of any resolved stars, still contains instrumental 
straylight from off-axis stars outside the aperture.

\subsubsection{Measurement of straylight profile of a star}

The stray radiation profile of a star, $I_{\rm stray}(r)$, in the range of 
$r\approx1\arcmin - 1\degr$ 
is thought to be caused mainly by scattering from the telescope main mirror 
micro-ripple and dust contamination (see e.g. \citealt{Beckers}).

Sirius was used as light source and the straylight was 
measured over the angular offset range of  $r$ = 89\arcsec--2000\arcsec\,.  
OFF positions were located at 1\degr\, North and 0.8\degr\, NE. The airmass was 1.1 - 1.2.
The closely linear form of log\,$I_{\rm stray}(r)$ vs. log\,$r$ enabled 
a safe fitting over 200\arcsec - 3000\arcsec\,. 
The straylight intensities were divided by the Sirius fluxes, adopted from \citet{Kur}, 
at the corresponding centre wavelengths of the five filters. They were corrected for 
the atmospheric extinction to correspond to the 'below the atmosphere' values of the straylight. 

For the 5 bands we have obtained the following fits:\newline
log\,$I_{\rm stray}$= -1.553$\pm0.004$\,log $r\arcsec\, -4.42\pm0.12$; for $u$\newline
log\,$I_{\rm stray}$= -1.484$\pm0.004$\,log $r\arcsec\, -4.54\pm0.13$; for 384/416/$b$\newline
log\,$I_{\rm stray}$= -1.538$\pm0.008$\,log $r\arcsec\, -4.57\pm0.25$; for $y$ 

For the Tycho $B_T$ and $V_T$ magnitudes we adopted the relations 
obtained for 384/416/470~nm and 555~nm, respectively. 
Compared with the 'canonical' (photographic) \citet{King} profile  
our $I_{\rm stray}$  levels at 10\arcmin\, offset are by  
a factor of $\sim$1.30 lower, and our slopes are shallower than that (2.0) of King.

The straylight vs angular distance profiles were applied to 
complete sets of stars adopted from comprehensive star catalogs  
to calculate the straylight in 5 colours for each target position.

\subsubsection{Straylight correction for the observed positions}

The  straylight intensities were calculated in the  $B_T$ and $V_T$ bands 
for each target position in the LDN\,1642 cloud area. 
For stars with $B_T \le$ 11.5~mag we adopted $B_T$ and $V_T$~magnitudes from
the Tycho-2 catalogue\footnote{http://archive.eso.org/skycat/servers/ASTROM}
\citep{Hog} up to a radius of 180~arcmin\,; for the fainter stars with $B > 11.5$~mag 
we adopted the $B$ and $V$ magnitudes of The Naval Observatory Merged Astrometric 
Dataset (NOMAD) catalogue\footnote{http://www.nofs.navy.mil/nomad/} 
(\citealt{Nomad04, Nomad05}) and calculated their straylight contributions 
up to a radius of 50~arcmin\,. 

The straylight from Tycho-2 stars was  $\sim0.3 - 1$ and $\sim0.4 - 1.5$\,S$_{10}$
for the bands $B_T$ and $V_T$, respectively.
The range  for the Nomad stars was $\sim0.14 - 0.40$,  $\sim0.14 - 0.41$ and 
$\sim0.3 - 1.0$ S$_{10}$ in $B$, $V$ and $R$, respectively.  

Our 35 positions were selected using the SRC Southern Sky Survey plates
 to exclude all stars above the limiting magnitude of
$R \approx 20^{\rm m}$. In order to check whether any fainter stars are found inside 
the \diameter86.55\arcsec photometer diaphragm area, we have used the GAIA  data base.
Because of possible slight setting errors or
(unlikely) drifting of the telescope positioning during the exposure, we have 
also looked for stars in a $10\arcsec$\, annulus around the diaphragm, that is within
$r \le 53.3\arcsec$ from the centre. In a few cases faint stars were detected that 
could have contributed mostly no more that 0.5  S$_{10}$ to the surface brightness. 

The straylight consists of the sum
of contributions from a large number of individual stars.
In no case was it caused by a single or a few stars near the observed position.
Thus, we have adopted the spectral energy distribution (SED) of the integrated 
starlight (ISL) at galactic latitude $|b|= 37\degr$ \citep{mat18} to calculate
the transformation factors from $B, V$ and $R$ to our five photometric bands.

The differential straylight corrections, Pos\,N minus Standard Pos\,8, were typically in the 
range $\la 0.5, 0.5, 0.7, 1$ and $0.5$ \cgs\, in $u, 384, 416, b$ and $y$, respectively.

\begin{table*}
\begin{minipage}{176mm}
 \caption{Estimation of H$_{\beta}$ line intensity from ionized gas at distances beyond LDN~1642. 
Intensities are in units of \cgs\, and refer to the H$_{\beta}$ line contribution in a filter 
band of 27 \AA\, width. Column (2) gives the observed differential H$_{\beta}$ intensity reltative to the opaque
standard position 8.}
\begin{tabular}{lcccccc}
Area & $\Delta I_{\rm H\beta}$(tot) & $I_b$(sca)       &$I_{\rm H\beta}$(sca)& $\Delta I_{\rm H\beta}$(sca) & $\tau_{\rm H\beta}$ & $I_{\rm H\beta}$(bcg)\footnote{$ = \exp(\tau_{\rm H\beta}) \times [\Delta I_{\rm H\beta}(tot)- \Delta I_{\rm H\beta}(sca)$] }\\ 
     &                             &              &$= 0.148\times I_b$(sca)       & rel. to Pos~8  &               & \\
(1)     & (2)                         &    (3)       & (4)                       & (5)            & (6)           &    (7)    \\
\hline
Centre\footnote{area includes positions 0, 1, 5, 6, 7, 8, 9, 10, 11, 12, 13, 40, 42, 43} &   4.2$\pm$0.1\footnote{Derived from differential spectrum Pos\,9/42 minus Pos\,8 \citep{mat17b}}
                   & 28.7$\pm$1.0 &4.25    &0.0                   & 0.90$\pm 0.02$ &10.3$\pm$0.3  \\

North\footnote{Positions 14,15,16,17,18,19,39,48} 
      &  9.7$\pm$0.7   &  7.0$\pm$0.7   &1.04              &-3.21                  &0.25$\pm0.02$  &16.5$\pm$1.2  \\

South\footnote{Positions 20,21,22,24,25,26,32} 
      & 3.9$\pm$0.8  &  5.8$\pm$1.1   & 0.86               &-3.39                  &0.20$\pm0.05$   & 8.9$\pm$1.2  \\

East\footnote{Position 3}    
& 6.5$\pm$3.0  & 18.2$\pm$1.7          & 2.69              &-1.56                  &0.53$\pm0.02$   &13.8$\pm$6.3  \\

West\footnote{Positions 34,35,36}  
      & 5.9$\pm$1.4 & 5.53$\pm$0.8&      0.81              & -3.44                 &0.16$\pm$0.12   &10.9$\pm$2.4  \\

\hline
\end{tabular}
\vspace{-5mm}
\end{minipage}
\end{table*}

\begin{table*}
\begin{minipage}{130mm}
 \caption{Photometry of Arc~A filament in the Eridanus superbubble. Positions 203 and 206 are on-source, 
201, 209 and 211 off-source. All surface brightnesses are in units of \cgs\,; for $H\beta$ see text.  
The extinctions are in magnitudes.}
\begin{tabular}{lcccccc}
\hline
                    & (2)            & (3)          & (4)          & (5)        & (6)                 & (7) \\
   Filter          & Pos 203\footnote{{\em Coordinates (J2000):} Pos\,203 04:01:39.9.0 +02:19:08; Pos\,206 04:02:38.4 +02:46:46; \newline Off\,201 03:56:06.4 +02:28:20; Off\,209 04:07:4.2 +01:44:01; Off\,211 04:07:55.1 +02:40:25}         & Pos 206$^{a}$        & Mean 203,206       &$A_{\lambda}$&$A_{\lambda}$-$A_{\beta}$& Mean,de-extincted\\
                   & Off\,201, Off\,209 &  Off\,211   &norm. $I_{\rm H\beta}=100$&            &                     &norm. $I_{\rm H\beta}=100$ \\
\hline
u                   & 17.6$\pm$1.8   & 11.2$\pm$3.2  &16.2$\pm$1.6  &1.59& 0.43  &24.1$\pm$2.4\\
384                 &  6.4$\pm$2.1   &  6.1$\pm$3.3  & 6.3$\pm$1.8  &1.51& 0.35  & 8.8$\pm$2.4\\
416                 &  7.0$\pm$2.2   &  3.8$\pm$3.2  & 6.3$\pm$1.9  &1.41& 0.25  & 7.9$\pm$2.4\\
b                   &  3.6$\pm$1.4   &  4.6$\pm$2.7  & 3.7$\pm$1.2  &1.23& 0.07  & 3.9$\pm$1.3\\
y                   &  4.4$\pm$2.4   &  1.4$\pm$7.1  & 4.0$\pm$2.2  &1.00&-0.16  & 3.5$\pm$1.8\\
$H\beta$            &106.4$\pm$1.5   & 76.7$\pm$5.5 &100.0$\pm$1.4  &1.16& 0.00 &100.0$\pm$1.4\\
\hline
\end{tabular}
\vspace{-5mm}
\end{minipage}

\end{table*}

\subsection{Correction for ionized gas emission}

Ionized gas along the line of sight contributes via line and, to a smaller part, also via continuum emission 
to the signal in all five filter bands. Its presence in the background radiation all over the sky is well 
demonstrated by the  H$\alpha$ emission line, see e.g. \citealt{reynolds}. Behind LDN\,1642
there is an enhanced background from the Orion-Eridanus bubble. Besides the 'direct' emission 
from ionized gas along the line of sight in front, $I_{\rm em }$(fgr), and behind of LDN\,1642, 
 $I_{\rm em }$(bgr), the ionized gas emission contributes also indirectly, via scattering off dust grains 
\citep{mat07,wood,witt10}.
The total ionized gas emission towards a position in the  LDN\,1642 area can be expressed as
\begin{equation}
I_{\rm em }{\rm (tot)} = I_{\rm em }{(\rm bgr)} e^{-\tau} + I_{\rm em }{(\rm fgr)} +  I_{\rm em}{(\rm sca)}
\end{equation}
where $\tau$ is the optical depth of dust through the cloud. 

\subsubsection{Measurements of H$\beta$ emission line}

As measure of the ionized gas emission we have used  the H$\beta$ emission line. We have measured the
line intensity through a narrow band filter (ESO\,326, $\lambda_0 = 4866$~\AA\,, $\Delta\lambda = 27$ \AA\,), 
and used the Str\"omgren $b$ band filter (ESO\,324, $\lambda_0 = 4673$ \AA\,, 
$\Delta\lambda$ = 169 \AA\,)
as reference filter. For each position N, the surface brightness differences
relative to the standard position\,8 were measured in both filters:
 \[\Delta I_{\rm obs}^{\rm Pos\,N}(4866)= I_{\rm obs}^{\rm Pos\,N}(4866)- I_{\rm obs}^{\rm Pos\,8}(4866)\] 
\[\Delta I_{\rm obs}^{\rm Pos\,N}(b)= I_{\rm obs}^{\rm Pos\,N}(b)- I_{\rm obs}^{\rm Pos\,8}(b)\]

Then, the H$\beta$ emission line intensities, relative to Pos. 8, were calculated from  
\[\Delta I_{\rm H\beta}^{\rm Pos\,N}(tot)=  \Delta I_{\rm obs}^{\rm Pos\,N}(4866) - \Delta I_{\rm obs}^{\rm Pos\,N}(b)\] 

Because of the relatively large scatter of the $\Delta I_{\rm H\beta}(tot)$ values
for individual positions we have defined five larger areas,
Centre, North, South, East and  West, see Table A1. 
The positions included into each of these areas are listed in the footnotes of the table.                          
The distribution of ionized gas emission in the LDN~1642 area is relatively smooth (see 
 \citealt{mat07}). It is, therefore, justified to use the mean values of  
$\Delta I_{H \beta}$(tot) as an approximation for the individual positions in each of the five areas.
The mean values of  $\Delta I_{\rm H\beta}(tot)$ for each
of these five areas are given in column (2) of the table.

\subsubsection{Correcting H$\beta$ emission for dust-scattered component}
 
In the direction of LDN\,1642 most of the H$\beta$ emission comes from the ionized  {\em background} gas
in the Ori-Eri bubble. The ionized {\em foreground} gas 
is assumed to be evenly distributed over the cloud area; its contribution has thus been omitted 
in the differential measurements of the present study. 
The observed  intensities  $\Delta I_{H\beta}$(tot) contain also the component $I_{\rm H\beta}$(sca), 
which is the H$\beta$ light {\em scattered} off the dust grains in the cloud, see equation (A9); 
for the purpose of deriving the LOS background component, $I_{\rm H\beta}$(bgr), a correction is needed for it.  
The scattered H$\beta$ line contribution can be estimated using the $b$ band scattered light intensity.

{\em In the Centre area}, instead of using narrow/broad band filter photometry, 
we have adopted the value of  $\Delta I_{\rm H\beta}$(tot) from \citet{mat17b} (see their Fig. 1)  
where it was determined using the difference spectrum between the translucent positions 9 and 42 
($A_{\rm H\beta}=0.98$\,mag and the opaque standard position 8. 
The scattered light intensities for Pos~9/42 and Pos~8 in the $b$ band are closely the same 
(see Fig.~6 in \citealt{mat17a}); the same is true also for the scattered H$\beta$ line contributions
and, therefore, the value of $\Delta I_{\rm H\beta}$(tot) does not need a scattered light correction.
 
{\em In the North, South, East and West areas} the intensity of the scattered  H${\beta}$ line is taken to be 
$I_{\rm H\beta}$(sca) = $f \times I_b$(sca). The proportionality factor $f$ is derived as follows:
Using a large spectroscopic data set obtained from the SDSS background positions,
 \citet{brandt} have derived a spectrum of the Diffuse Galactic (scattered) Light (DGL).
This spectrum represents at the same time the incident Galactic background light spectrum.
The Balmer lines  H$\beta$ and  H$\alpha$ appear as emission lines. 
The observed H$\beta$ emission line equivalent width is  4.0$\pm0.7$ \AA\,.
For the standard position 8 the continuum intensity near H$\beta$ is
$I_{\rm cont}(4861 \AA\,) \approx I_{\rm cont}(b) = 28.7$ \cgs\,.  For  
$\Delta\lambda$ = 27 \AA\, filter  the equivalent width of 4.0 \AA\, corresponds to an intensity 
of  $I(H\beta$) = $4.0/27 \times I_{\rm cont} = 0.148 \times I_{\rm cont}= 4.25$ \cgs\,. 
For observed positions with different amounts of scattered continuum light, $I_b$(sca),
 the H$\beta$ scattered light intensity varies accordingly, that is 
$I_{\rm H\beta}$(sca) = $0.148 \times I_b$(sca). These values are given in column (4) of Table A1.

According to equation (A9) the observed H$\beta$ line intensities for Area\,A and Pos\,8 are given by\newline

\noindent $I^{\rm A}_{\rm H\beta}({\rm tot}) = I^{\rm A}_{\rm H\beta}{(\rm bgr)} e^{-\tau(A)} + I^{\rm A}_{\rm H\beta}{(\rm fgr)} +  I^{\rm A}_{\rm H\beta}({\rm sca)}$,  and \newline

\noindent $I^{\rm 8}_{\rm H\beta}({\rm tot}) = I^{\rm 8}_{\rm H\beta}{(\rm bgr)} e^{-\tau(8)} + I^{\rm 8}_{\rm H\beta}{(\rm fgr)} +  I^{\rm 8}_{\rm sca}{(\rm sca)}$ \newline

Because  $e^{-\tau(8)} \approx 0$ and 
$I^{\rm A}_{\rm H\beta}{(\rm fgr)}\approx  I^{\rm 8}_{\rm H\beta}{(\rm fgr)}$,
the  H$\beta$ background intensity for Area\,A, corrected for scattered light, is given by:
\[I^{\rm A}_{\rm H\beta}({\rm bcg})= \]
\[\quad\, =e^{\rm \tau(A)}\left\{I^{\rm A}_{\rm H\beta}({\rm tot})-I^{\rm 8}_{\rm H\beta}({\rm tot})-[I^{\rm A}_{\rm H\beta}(\rm sca)-I^8_{\rm H\beta}(\rm sca)]\right\}\]
The values 
$I^{\rm A}_{\rm H\beta}(\rm sca)-I^8_{\rm H\beta}(\rm sca)= \Delta I^{\rm A}_{\rm H\beta}(\rm sca)$ and $I^{\rm A}_{\rm H\beta}(\rm bcg)$
are given in columns (5) and (7) of Table A1.
As can be seen from Table~A1 there is a substantial gradient in the H$\beta$ background emission,
with increasing values from South 
to North. This same behaviour is seen also in the H${\alpha}$ intensities \citep{mat07}:
$I_{\rm H\alpha}$ in the Northern area is ca. twice the value in the South.

\subsubsection{Subtraction of the ionized-gas contribution from the 5-colour photometry bands}

Emission by the ionized background gas contributes to the surface brightness signal in
all five bands of our photometry. 
The contributions are proportional to the $I_{\rm H\beta}$(bcg) values as given in 
column (7) of Table A1. In order to determine the proportionality factors
we have observed a bright  filament, called Arc A, in the Eridanus superbubble area, 
$\sim$18\degr distant from LDN~1642. (see e.g. \citealt{boumis}; \citealt{ochsendorf}). 
In this filament $I_{\rm H\beta} \sim 100$ \cgs\,,
that is $\sim10$ times brighter than in the LDN~1642 area; this enabled a measurement dominated
by the ionized gas emission, even though some dust scattered light may be present.
The observations were made using the same equipment, 
differential observing method, and way of analysis as for LDN~1642 cloud area. 
The observed intensities are given in Table A2. The observed lines of sight have modest 
extinction caused by the foreground translucent cloud LDN~1569. We give in column (7)
the extinction-corrected intrinsic intensities, normalized to 
$I_{\rm H\beta} = 100$~\cgs\,.

The ionized-gas corrections are now done as follows:
the  $I_{\rm H\beta}({\rm bg})$ background intensities, $r > r({\rm LDN~1642}$), are assumed 
to be constant in each of the areas, Centre, North, South, East and West;
the effect of the foreground extinction 
is corrected individually for each position and each filter band. 
The resulting ionized-gas corrections are small, remaining below $\le$ 3.1, 1.1, 1.0, 0,5, 0.5 \cgs\, 
in the filters u, 384\,nm, 416\,nm, b, and y, respectively.

\subsection{Correction for zodiacal light gradient}

 The zodiacal light shows a substantial gradient over the angular extension, $4.5\degr \times 4\degr$, 
 of our target area. The area is centered at the ecliptic 
coordinates $\lambda$ = 64.5$\degr$, $\beta$ =  -35.8$\degr$ (J2000).
The helioecliptic longitude of the area was in the range
 $\lambda - \lambda_{\sun} \approx 160\degr - 175$$\degr$ and varied 
substantially during an observing period of a week.
The zodiacal light gradients over the latitude and longitude ranges 
were $\sim$5 and $\la 1$\,\,\cgs\,, respectively. 

The outermost positions in the LDN~1642 outskirts are almost free of scattered light from dust.
In addition, a number of positions have only small or moderate scattered-light 
contributions and can be utilized for the ZL gradient removal. Their far-IR emission is used as
measure of the dust content. All positions in the target area 
have been observed at 200 $\mu$m using the absolute photometry mode 
of the ISOPHOT instrument aboard the  {\em Infrared Space Observatory ISO} (see Section 2.3) .
At small and moderate optical depths with $A_{\lambda}\la1$~mag, corresponding to 
$I_{200}\la22$~MJy~sr$^{-1}$ in $b$ and $y$, and $\la15$~MJy~sr$^{-1}$ in $u$/384/416~nm, 
a linear relationship between these quantities has been found, see Section 2.3 and Fig.\,2. 

\begin{table*}
\begin{minipage}{105mm}
 \caption{The coefficients A, B, C, and D  in the five intermediate-band filters, 350~nm(u), 384~nm, 
416~nm, 470~nm(b) and 550~nm(y) as obtained from the least-squares fit according to equation (9).
The unit is \cgs\,/degr for A and B,  \cgs\,/MJy~sr$^{-1}$ for C and  \cgs\, for D and the rms.}
\begin{tabular}{lccccc}
Filter              & A     & B       & C       & D & rms      \\
                    & (2)   & (3)     & (4)     &    (5) & (6)  \\
\hline
u                   & 0.48$\pm$0.45 & 1.83$\pm$0.29 & 1.27$\pm$0.22 & -8.12$\pm$1.37 & 1.818  \\
384                 &-0.39$\pm$0.32 & 1.24$\pm$0.20 & 1.75$\pm$0.14 &-15.65$\pm$0.89 & 1.288  \\
416                 & 0.16$\pm$0.45 & 1.56$\pm$0.31 & 2.03$\pm$0.16 &-22.60$\pm$1.22 & 2.081  \\
b                   & 0.58$\pm$0.44 & 1.56$\pm$0.35 & 1.68$\pm$0.11 &-28.59$\pm$1.03 & 2.575  \\
y                   & 0.30$\pm$0.51 & 1.88$\pm$0.37 & 1.40$\pm$0.08 &-31.85$\pm$0.95 & 2.280  \\
\hline
\end{tabular}
\end{minipage}
\end{table*}

The observed differential surface brightnesses in this paper have so far been referred 
to a zero point as defined by Pos~8, $I_{\rm opt}$(Pos\,8)= 0.
We want to refer them to a zero level corresponding to zero dust
column density. The absolute photometry at 200~$\mu$m, corrected for CIRB and ZL, makes this
approach possible, see Section 2.3. For this pupose, and also for the purpose of subtracting the ZL foreground 
gradient over the observed area, we have made a least-squares fit to the surface brightnesses 
$I_{\rm opt}$ for each of the five optical filters of the following form:
\begin{equation}
I_{\rm opt} = Ax + By +Cz + D 
\end{equation}
The two spatial coordinates are $x = \lambda - \lambda_0$ and $y = \beta - \beta_0$ 
where $\lambda_0$ and $\beta_0$ are the coordinates of the standard position, Pos~8;
the third coordinate is the far-IR intensity $z = I_{200}$.
A large number of OFF positions with almost none or only little dust were selected.
These positions were complemented with a few positions with $0.5\la A_{\lambda}\la1$~mag
in order to improve the accuracy of the coefficient C and that of the zero
point D. The number of positions used for the fit was 23 for $u$ and 384,
25 for 416, 28 for $b$ and 31 for $y$. 
The resulting values and standard errors of the coefficients $A, B, C$ and $D$, 
as well as the mean rms of the fitted data points are given in Table\,A3.

The zero point and the ZL foreground gradient correction are thus given by 
\begin{equation}
\Delta I_{\rm opt} = Ax + By + D 
\end{equation}
These $\Delta I_{\rm opt}$ corrections are subtracted from
 the five intensities, $I_{\rm opt}(u)$,  $I_{\rm opt}(384)$,  $I_{\rm opt}(416)$,  $I_{\rm opt}(b)$  
and $I_{\rm opt}(y)$ at each observed position $x, y$.
The corrected values for the standard position Pos\,8, located at $x = y = 0$, will thus be 
$ I_{\rm opt}$(Pos\,8) = $-D$.\\

\section{Extinction in the centre of LDN~1642 from NTT/SOFI $HK_s$ photometry}

Positions 8 and 40, in the centre of the cloud have no 2MASS
stars within the  $\O$$88\arcsec$ surface-photometry aperture area and not even within $\O$$2\arcmin$.
For these positions we have made use of NTT/SOFI H and K$_s$ band imaging data (Programme 077.C-0338(A), 
observing dates 2006-08-11 - 2006-08-13). Using the
colour excesses $E(H-K_s)$ of 11 stars located within a $2\arcmin \times 4\arcmin$  area 
encompassing both these  closeby ($\sim 2.5\arcmin$ apart) positions we found a mean extinction 
of $A_V = 16.1 \pm 1.4$\,mag. As for the 2MASS data we used the conversion ratio 
$A_V/E(H-K_s) = 15.98$ corresponding to $R_V = 3.1$. 

\end{document}